\documentclass[aps,showpacs,pra,twocolumn,citeautoscript,superscriptaddress,floatfix,longbibliography]{revtex4-1}

\usepackage{amsthm}
\usepackage[utf8]{inputenc}
\usepackage{amsmath,amssymb,amsmath,amsthm,bm,graphicx,xcolor}
\usepackage{natbib}
\usepackage[breaklinks]{hyperref}
\usepackage{color}
\usepackage{tabularx}
\usepackage{epsfig}
\usepackage{amssymb}
\usepackage{graphicx}
\usepackage{wasysym}
\usepackage{dcolumn}
\usepackage{epstopdf}
\usepackage{subfigure}
\usepackage{psfrag}
\usepackage{bbold}
\usepackage[breaklinks]{hyperref}
\hypersetup{colorlinks=true}
\usepackage{physics}

\newcommand{\KS}{{\rm{KS}}}

\newcommand{\gs}{\rm gs}

\newcommand{\vecform}{\bm} 
\newcommand{\xx}{\vecform{x}}

\newcommand{\rr}{\bm{r}}         
   
\newtheorem{thm}{Theorem}

\newcommand{\nk}{{n\bm{k}}}
\newcommand{\mint}[2]{\int\!\!{\rm d}^{#1}{#2}\:}
\newcommand{\mdint}[3]{\mint{#1}{#2}\!\!\!\mint{#1}{#3}}

\begin{document}

\title{Reduced Density Matrix Functional Theory for Superconductors}

\author{Jonathan Schmidt}
\affiliation{Institut f\"ur Physik, Martin-Luther-Universit\"at
Halle-Wittenberg, 06120 Halle (Saale), Germany}

\author{Carlos L. Benavides-Riveros}
\email{carlos.benavides-riveros@physik.uni-halle.de}
\affiliation{Institut f\"ur Physik, Martin-Luther-Universit\"at
Halle-Wittenberg, 06120 Halle (Saale), Germany}

\author{Miguel A. L. Marques}
\affiliation{Institut f\"ur Physik, Martin-Luther-Universit\"at
Halle-Wittenberg, 06120 Halle (Saale), Germany}

\date{\today}

\begin{abstract}
We present an \textit{ab initio} theory for superconductors, based on
a unique mapping between the statistical density operator at
equilibrium, on the one hand, and the corresponding one-body reduced
density matrix $\gamma$ and the anomalous density $\chi$, on the
other. This new formalism for superconductivity yields the existence
of a universal functional $\mathfrak{F}_\beta[\gamma,\chi]$ for the superconductor
ground state, whose unique properties we derive. We then prove the
existence of a Kohn-Sham system at finite temperature and derive the
corresponding Bogoliubov-de Gennes-like single particle equations. By
adapting the decoupling approximation from density functional theory
for superconductors we bring these equations into a computationally
feasible form. Finally, we use the existence of the Kohn-Sham system
to extend the Sham-Schl\"uter connection and derive a first
exchange-correlation functional for our theory. This reduced density
matrix functional theory for superconductors has the potential of
overcoming some of the shortcomings and fundamental limitations of
density functional theory of superconductivity.
\end{abstract}

\maketitle

\section{Introduction}

Superconductivity was discovered more than one 
cen\-tury ago, when Kamerlingh Onnes observed, 
and wrote in his lab notes, that ``mercury’s 
resistance is practically zero'' at very low temperatures 
\cite{PhysToday}. While the first phenomenological 
theory of superconductivity was developed 24 years 
later in the form of the London equations \cite{LONDON, deGennes}, 
it took nearly half a century until the first 
successful microscopic description emerged~\cite{PhysRev.108.1175}. 
The so-called Bardeen-Cooper-Schrieffer (BCS) theory, based on the concept 
of Cooper pairs and a non-particle number conserving 
Ansatz for the wave function, predicted a pair-bound-state 
in the presence of an attractive force, regardless 
of the strength of the interaction. Although BCS theory 
is able to explain some universal features of superconductors 
(such as the ratio between the energy gap and the critical 
temperature), a proper treatment of the strong electron-phonon 
coupling regime had to wait until the theory developed 
by Eliashberg \cite{osti_7354388}. 

It is by and large accepted that the electron-phonon interaction 
is well accounted for within BCS and Eliashberg theories. Yet 
the correlation effects due to the electron-electron Coulomb 
repulsion are extraordinarily difficult to handle, and they are 
usually compressed in an adjustable parameter (namely, $\mu^*$). 
Since the adjustability of the parameter diminishes the 
predictive capability of the theory, it comes as no 
surprise that the correct description of superconductors 
remains one of the great questions in condensed 
matter theory. We still have a poor understanding of the 
superconducting features of unconventional superconductors, 
like high-$T_c$ cuprates 
\cite{bednorz1986possible,RevModPhys.78.17,PhysRevLett.119.237001} or layered 
organic materials \cite{0034-4885-74-5-056501}. Moreover, 
recent discoveries of superconductivity in two layers of 
graphene \cite{graphene,PhysRevLett.121.087001},
and in compressed hydrides at extreme 
pressure~\cite{Eremets,PhysRevLett.114.157004,PhysRevLett.122.027001},
continue to challenge our understanding of superconducting systems.

Proposed by Oli\-veira, Gross and Kohn in 1988, density functional 
theory for the superconducting state (SC-DFT) provided an \textit{ab initio}
 unified treatment of correlation and inhomogeneity effects in superconductors~\cite{PhysRevLett.60.2430}.  Extending the famous Hohenberg-Kohn 
theorems of normal density functional theory (DFT) \cite{HK}, SC-DFT  
states a one-to-one mapping 
between the equilibrium statistical density operator 
$\hat\varrho_{\rm eq}$ and the corresponding electronic 
$n(\rr) \equiv \sum_\sigma \langle
\psi_\sigma^\dagger(\rr) \psi_\sigma(\rr)\rangle$ and anomalous
$\chi(\rr,\rr')\equiv \langle \psi_{\uparrow}(\rr)
\psi_{\downarrow}(\rr')\rangle$ densities:
\begin{align}
  \label{eq0}
  (n,\chi) \overset{1-1}{\longleftrightarrow}   \hat\varrho_{\rm eq}
  \equiv \frac{e^{-\beta (\hat{H}_{v,\Delta}-\mu\hat N)}}{{\Tr}[e^{-\beta(
      \hat{H}_{v,\Delta} - \mu \hat N)}]},
\end{align}
with $\hat{H}_{v,\Delta}$ being the Hamiltonian for a superconductor
in an external potential $v(\rr)$ and a non-local pairing potential
$\Delta(\rr,\rr')$. This latter term is included to break the $U(1)$
symmetry in the superconducting phase. If not otherwise indicated,
``$\Tr$'' always means the trace in Fock space.  Since SC-DFT accounts for
electron-electron interactions in the same way as normal DFT, there is
no use of any semi-empirical parameter. Hence, SC-DFT is a truly
\textit{ab initio} theory for superconductors.

Formally, SC-DFT is able to describe superconductivity in
all systems, for the many-electron problem is cast into a universal
exchange-co\-rre\-lation functional whose existence is ensured by the
Oliveira-Gross-Kohn theorem. Further work on this
topic~\cite{SCDFT1,SCDFT2} extended SC-DFT to include the nuclear
density as a third component. Recently, the theory was revisited to
allow the inclusion of magnetic
fields~\cite{PhysRevB.92.024505,PhysRevB.92.024506}.

SC-DFT has been extremely successful in predicting superconductivity
in a wide variety of materials
\cite{cac6scdft,dissilanescdftpressure,PBSCDFT}, and in particular in
high-pressure research
\cite{metalsscdftpressure,sulfurhybridesscdftpressure,seleniumscdftpressure}.
Yet, the practical applicability of SC-DFT rests (as in the standard
DFT) upon a Kohn-Sham scheme, which maps the real interacting system
of interest to an auxiliary noninteracting one with the same
equilibrium electronic and anomalous densities $(n,\chi)$. The
existence of such a Kohn-Sham system is problematic, however. In fact,
in a recent paper we proved that such a noninteracting system does not
exist, at least at zero temperature \cite{firstpaper}. Furthermore,
there is an unpleasant asymmetry in SC-DFT that complicates
considerably certain key points in the derivations. This asymmetry is
related to the use of the local electron density [there is only one
  $\rr$ in $n(\rr)$] and the non-local anomalous density [there are
  two different $\rr$ and $\rr'$ in $\chi(\rr,\rr')$]. Although no
problem arises at the level of the one-to-one correspondence of
Eq.~\eqref{eq0}, we can expect complications when developing
exchange-correlation functionals for SC-DFT. Indeed, for
superconductivity only a few functionals of the electronic and
anomalous densities have been developed.  Even worse, these
functionals require completely \textit{ad hoc}, and rather arbitrary,
modifications in order to yield good results \cite{boeri1}.

It is a well-know fact that the most spectacular fai\-lu\-res of
normal DFT (with standard exchange-correlation functionals) are
related to an incorrect description of strongly (static, in
quantum-chemistry jargon) correlated systems
\cite{Cohen792,Sanchez}. Since Kohn-Sham DFT does not deal with
fractional occupation numbers, the electronic structure of
multireference systems is still an open problem within the theory.
Reduced density matrix functional theory (RDMFT) is a natural
extension of DFT, whose aim is to exploit the one-electron picture of
the many-body wave function $\ket{\Psi}$ by seeking a functional of
the one-electron reduced density matrix $\hat \gamma = N{\Tr}_{N-1}
[\ket{\Psi}\bra{\Psi}]$, allowing therefore fractional occupation
numbers \cite{Gilbert,Schade2017}.  RDMFT describes closed-shell
molecular systems with accuracies higher by one order of magnitude
than DFT \cite{ML, PirisPRL}. It has also succeeded in predicting more
accurate gaps of conventional semiconductors \cite{LSDEMG}.
Furthermore, RDMFT correctly captures the physics of the
insulator-metal phase transition of transition metal oxides
\cite{Sharma}, something, as is well known, DFT cannot do.

In this article, we develop a new \textit{ab initio} theory of
superconductivity, namely, a RDMFT formalism for superconductivity
(from now on, SC-RDMFT). This is done by using $\gamma$ instead of $n$
as one of the basic variables of the theory.  In this way, we expect
that most of the advantages of RDMFT will translate well to strongly
correlated superconducting systems. Furthermore, the use of the density-matrix restores the symmetry of the equations, and as we will see,
simplifies considerably some derivations.

We then derive  the existence of a Kohn-Sham system which results 
in a set of coupled Bogoliubov-de-Gennes-like
equations. These equations are still extremely difficult to solve as
they require precision on the level of the superconductive coupling
while encompassing the energy scale of the whole band structure \cite{Zhu}. In
order to simplify the Kohn-Sham equations into a form that can
feasibly be solved we appropriate the so-called decoupling
approximation from SC-DFT. It allows us to separate the solution of
the standard electronic problem, the influence of the lattice and the
superconducting coupling. Analogously to DFT, the challenges of the
many-body problem cannot be avoided through a Kohn-Sham system but
only be transferred to the exchange-correlation functional. We suggest
a first solution to this challenge by introducing the Sham-Schl\"uter
connection \cite{ss1,ss2,ss3} to SC-RDMFT and deriving a first
functional. Moreover, the derivation avoids several ad-hoc
approximations that are required in SC-DFT.

The paper is organized as follows. The first section is this
introduction.  For the paper's completeness, Sec.~\ref{introRDMFT}
summarizes normal RDMFT and its advantages in the light of normal DFT.
In Sec.~\ref{section3} we present the generalized one-body reduce
density matrix in Nambu-Gorkov space. Such a matrix is the correct
variable to deal with the spon\-ta\-neous breaking of the $U(1)$
symmetry in the superconducting phase. Section~\ref{section4} presents
the theoretical foundations of SC-RDMFT.  
We derive a Gilbert theorem
and state the mathematical properties of the universal
exchange-correlation functional for superconductors.  In
Sec.~\ref{KohnSham} we prove the existence of the Kohn-Sham system at
finite temperature, followed by the decoupling approximation in
Sec.~\ref{decoupling} and the Sham-Schl\"uter connection in
Sec.~\ref{SSC}.  Section~\ref{conclusions} is devoted to the
conclusions of the paper.

\section{Reduced density functional theory in a nutshell}
\label{introRDMFT}

As a solution of the Schr\"odinger equation, the state $\ket{\Psi_{\gs}}$ 
describes the ground-state quantum system completely. The Hohenberg-Kohn 
theorems imply the existence of a universal functional of the external 
potential and the electronic density which reaches a minimum when evaluated on 
the ground-state electronic density. In 1975, Gilbert proved an extension of 
such theorems, showing that there is also a one-to-one correspondence between the 
ground-state wave function of a non-degenerate many-body system and the corresponding 
one-body reduced density matrix \cite{Gilbert}: 
\begin{align}
 \ket{\Psi_{\gs}} \overset{1-1}{\longleftrightarrow} \hat\gamma_{\gs}
\equiv N { \rm Tr}_{N-1}[\ket{\Psi_{\gs}}\langle\Psi_{\gs}|],
\label{eq:GILBERTone-to-one}
\end{align}
where $N-1$ particles are traced out.
The main advantage of this theorem is that any observable of the system 
(in its ground state) can be written as a functional of $\hat\gamma_{\gs}$. 
In practice, this latter condition is relaxed and the functional is evaluated
on the set of fermionic one-body reduced density matrices $\{\hat\gamma\}$.
More importantly, the functional of the kinetic energy,
which is unknown in DFT and should be included in the exchange-correlation 
functional, is known exactly in terms of $\hat\gamma$. In RDMFT, the
functional for the ground-state energy reads: 
\begin{align}
\mathcal{E}[\gamma] = \mdint{4}{x}{x'} h(\rr,\rr')  \gamma(\xx;\xx')
+ \mathcal{F}[\gamma],
\end{align}
where $\gamma(\xx;\xx') = \bra{\xx'}\hat \gamma \ket{\xx}$. 
The one-particle Hamiltonian 
$h(\rr,\rr') = -\tfrac12\delta(\rr-\rr')\nabla^2_{\rr} + v_{\rm ext}(\rr,\rr')$ 
contains the kinetic and external-potential operators. $\mathcal{F}[\gamma]$ is 
an unknown universal functional of the ground-state one-body reduced density 
matrix. 
We used the customary compact notation for spin and position 
coordinates $\xx \equiv (\rr, \sigma)$ (with $\sigma\in\{\uparrow, \downarrow\})$.

Another important advantage of Gilbert's theorem is that one can 
consider a broader set of nonlocal external potentials $v_{\rm ext} (\rr,\rr')$, which 
arise when the quantum problem is formulated in terms of many valence 
electrons, subject to an external potential of fixed nuclei and core 
electrons \cite{Gilbert}. Yet, the exact form of the exchange-correlation 
functional $\mathcal{F}[\gamma]$ is, by and large, not available and 
therefore the predicted RDMFT energy, with approximate functionals, can be either lower or higher than the exact 
ground-state energy \cite{MULLER1984446, Lieb,Benavides-Riveros2012,0953-8984-29-42-425602}.
Functionals in RDMFT are often engineered as approximate expressions 
of the two-body reduced density matrix
\begin{equation}
  \hat\gamma_2 \equiv \binom{N}{2} { \rm Tr}_{N-2}        
   [|\Psi\rangle\langle\Psi|] .
\end{equation}
This is normally accomplished by writing $\hat \gamma_2$ 
in terms of the ex\-chan\-ge-correlation hole $\rho^{\rm hole}_{\rm xc}(\xx,\xx')$, 
defined by the following relation:
\begin{align}
\gamma_2(\xx;\xx') \equiv \tfrac{1}{2} \gamma(\xx;\xx)\big[\gamma(\xx';\xx') -
    \rho^{\rm hole}_{\rm xc}(\xx,\xx')\big].
\end{align}
The electronic density, the main object in DFT, is of course the trace on 
spin space of the diagonal $\gamma(\xx, \xx)$. 

For example, the famous M\"uller (for historical reasons also
called Buijse-Baerends) functional describes the
exchange-cor\-re\-lation hole as the square of a \textit{hole
  amplitude} \cite{buijse}, reading
\begin{equation}
\label{eq:sem}
  \left|\frac{\gamma^{1/2}(\xx;\xx')}{\sqrt{\gamma(\xx;\xx)}}\right|^2,
\end{equation}
where $\hat \gamma^{1/2} \equiv \sum_i n^{1/2}_i \ket{\varphi_i} \bra{\varphi_i}$
is written in terms of the so-called natural occupation numbers
$\{n_i\}$ and natural orbitals $\{\varphi_i\}$, namely, the
eigenvalues and eigenvectors of $\hat \gamma$. Further developments in
RDMFT are inspired by this seminal functional
\cite{GU,CioPer,GPB,CGA, PhysRevB.66.155118}.  There is another perspective by studying
the cumulant part of $\gamma_2$ (i.e., $\gamma_2 - \tfrac12
\gamma \wedge \gamma$) under some of its known representability
conditions \cite{eltit, PhysRevLett.119.063002}.  Remarkably, almost
all RDMFT functionals fare quite well in benchmarking tests, yielding errors
for the correlation energy an order of magnitude smaller than
B3LYP, {\rm perhaps the most popular DFT functional in quantum chemistry,
and a  precision comparable to M{\o}ller-Plesset second-order perturbation   
theory~\cite{LM}}.
RDMFT has also succeeded in predicting more accurate gaps of conventional
semiconductors than semi-local DFT does. Furthermore it has demonstrated
insulating behavior for Mott-type insulators
\cite{Sharma,1367--17-9-0930382630,Pernalrev}. Recently, the role of
the generalized Pauli exclusion principle has been stressed within
RDMFT for pure states
\cite{progress,C7CP01137G,CSQ,RDMFTpvse,Schilling2017,Halle,PhysRevLett.122.013001, TDRDMFT}.

When researching superconductors we naturally want to consider their 
behaviour at finite temperature as their most important property is 
their transition temperature. Fortunately, a finite-temperature 
RDMFT was already developed by Baldsiefen et al.~\cite{PhysRevA.92.052514, PhysRevA.96.062508}.  
Just as in finite-temperature DFT, the considered systems are 
grand canonical ensembles and instead of the energy, for a Hamiltonian
$\hat H$, we consider the grand canonical potential $\Omega_{\mu,\beta}[\varrho]$
for fixed temperature $\beta^{-1}$ and chemical potential $\mu$:
\begin{equation}
\label{eqOmega}
\Omega_{\mu,\beta}\left[\varrho\right] =
\text{Tr} \hat \varrho\left(\hat{H}-\mu\hat{N}+\frac{1}{\beta}\ln\hat\varrho\right).
\end{equation}
The appropriate space for the statistical density operators is 
the Fock space. Instead  of the ground state, we are interested 
in the thermodynamic equilibrium. For general quantum systems, 
$\Omega_{\mu,\beta}\left[\varrho\right]$ is bounded from below by 
$\Omega_{\mu,\beta}\left[\varrho_{\rm eq}\right]$,
where $\hat\varrho_{\rm eq}$ is the equilibrium statistical density operator. 
In contrast to zero-temperature RDMFT, Gilbert's theorem in finite-temperature 
RDMFT is completely invertible: there is a one-to-one mapping between the 
equilibrium statistical density operator, the external potential 
(minus the chemical potential)  and the corresponding equilibrium one-body 
reduced density operator 
\cite{PhysRevA.92.052514,PhysRevB.82.205120}:
\begin{align}
\hat{v}_{\rm ext} - \mu \overset{1-1}{\longleftrightarrow} 
\hat\varrho_{\rm eq} \overset{1-1}{\longleftrightarrow} 
\hat\gamma_{\rm eq} \equiv 
N { \rm Tr}_{N-1}[\hat\varrho_{\rm eq}] .
\label{eq:GILBERTone-to-one2}
\end{align}
 The proof is analogue to Mermin's proof 
for finite-tem\-pe\-rature DFT~\cite{PhysRev.137.A1441}. 
This result implies that $\hat\varrho_{\rm eq}$
can be written as a functional of $\hat\gamma_{\rm eq}$, and therefore,
in the thermodynamic equilibrium, the functional \eqref{eqOmega} can be 
written as a functional of the equilibrium one-body reduced density matrix 
$\hat \gamma_{\rm eq}$, namely:
\begin{gather}
\Omega_{\mu,\beta}\left[\hat\gamma_{\rm eq}\right] = 
\text{Tr}\hat\varrho_{\rm eq}[\hat\gamma_{\rm eq}]\left(\hat{H}-
\mu\hat{N}+\frac{1}{\beta}\ln\hat\varrho_{\rm eq}[\hat\gamma_{\rm eq}]\right).
\end{gather}
It is worth saying that Eq.~\eqref{eq:GILBERTone-to-one2} is an 
advantage over the standard zero-temperature RDMFT. Indeed, there is no 
problem with degeneracy since degenerate states get the same 
equilibrium statistical density operator \cite{PhysRevB.82.205120}.

\section{Nambu-Gorkov one-body reduced density matrix}
\label{section3}

As presented in Section \ref{introRDMFT}, RDMFT is unfortunately not
sufficient for the description of superconducting systems.  The reason
for this lies in the spontaneous breaking of the $U(1)$ symmetry,
which implies that the particle number is not conserved.  The breaking
of such a symmetry also results in a finite expectation value of the
anomalous density $\chi(\rr,\rr')=
\langle\psi_{\uparrow}(\rr)\psi_\downarrow(\rr')\rangle$.  In
Nambu-Gorkov space, field operators are usually written as the
following spinors:
\begin{gather}
\bar{\Psi}_{\sigma}(\rr) =
\bigg(\begin{array}{c} 
\psi^{\dagger}_{\sigma}(\rr) \\ 
\psi_{-\sigma}(\rr) \end{array}\bigg).
\end{gather}
We write the generalized Nambu-Gorkov one-body reduced density 
matrix as the expected value of a product of Nambu-Gorkov field-operators, namely:
\begin{align}
\Gamma_{\sigma\sigma}(\rr,\rr') &=
\langle \bar{\Psi}_{\sigma}(\rr) \otimes
        \bar{\Psi}^\dagger_{\sigma}(\rr')\rangle .
\label{contraction}
\end{align}
Notice that $\Gamma_{\sigma\sigma}(\rr,\rr')$ can be denoted in terms 
of the reduced density matrix $\gamma_{\sigma\sigma}(\rr,\rr')$ and the anomalous 
density $\chi(\rr,\rr')$ as:
 \begin{align}
   \Gamma_{\sigma\sigma}(\rr,\rr') &= \begin{pmatrix}
  \langle \psi^\dagger_{\sigma}(\rr) 
          \psi_{\sigma}(\rr')\rangle &
  \langle \psi^\dagger_{\sigma}(\rr) 
          \psi^\dagger_{-\sigma}(\rr')\rangle \\
  \langle \psi_{-\sigma}(\rr) 
          \psi_{\sigma}(\rr')\rangle &
  \langle \psi_{-\sigma}(\rr) 
          \psi^\dagger_{-\sigma}(\rr')\rangle 
	\end{pmatrix} 
\label{eq:Gamma} \\ &= 
\begin{pmatrix}
	\gamma_{\sigma\sigma}(\rr,\rr') & \chi^{\dagger}(\rr,\rr') \\ 
	\chi(\rr,\rr') & \delta(\rr-\rr')-\gamma_{-\sigma-\sigma}(\rr,\rr')
\end{pmatrix}.
\nonumber
 \end{align}
Note also that the Nambu-Gorkov one-body reduced density ma\-trix is not 
properly normalized. It integrates to the volume, instead of integrating to the 
particle number. However, one can still obtain the average particle 
number from the ordinary one-body reduced density matrix $\gamma$ and the 
standard deviation of the particle number from the anomalous density $\chi$. 

Obviously this is only one of several possible choices. For instance,
in order to describe a system that is not symmetric with respect to 
spin direction one would have to consider the full Nambu-Gorkov one-body 
reduced density matrix in spin space:
\begin{gather}
{\bf \Gamma}(\rr,\rr') = \begin{pmatrix}
\Gamma_{\uparrow\uparrow}(\rr,\rr')&\Gamma_{\uparrow\downarrow}(\rr,\rr')
\\
\Gamma_{\downarrow\uparrow}(\rr,\rr')&\Gamma_{\downarrow\downarrow}(\rr,\rr')
\end{pmatrix}
\end{gather}
In this notation, the anomalous densities with parallel spin describe 
triplet Cooper pairs (experimentally verified eight years 
ago by Sprungmann et al.~\cite{triplet}). For this paper we will limit 
ourselves to systems with spin-rotational symmetry. Therefore, as a matter of simplification, we will omit the index $\sigma\sigma$ 
 from $\Gamma_{\sigma\sigma}$ from now on.

\section{SC-RDMFT: Theoretical foundations}
\label{section4}

This section is devoted to prove the fundamental theorems 
of our new SC-RDMFT, namely, a Gilbert theorem 
for systems with spontaneous breaking of the $U(1)$ symmetry. 
We formulate the variational principle in order to include 
ensemble $N$-representable $\Gamma$ in a 
formalism \textit{\`a la} Levy-Lieb. We then prove several 
properties of the universal functional that turn out to
 be crucial for minimization procedures and the development of 
SC-RDMFT functionals.

\subsection{Gilbert theorem for superconducting systems}

We will develop RDMFT for superconductors analogously to 
SC-DFT and accordingly we start with the Hamiltonian for
superconductors extended to non-local potentials:
\begin{widetext}
\begin{align}
\hat{H}_{v,\Delta} &= -\frac12\sum_\sigma
\mint{3}{r} \hat{\psi}^\dagger_\sigma(\rr) \nabla^2 \hat{\psi}_\sigma(\rr)   + \sum_{\sigma}
\mdint{3}{r}{r'} \hat{\psi}^\dagger_{\sigma}(\rr')v(\rr,\rr')\hat{\psi}_\sigma(\rr)
- \mdint{3}{r}{r'} \left[ \Delta^*(\rr,\rr')\hat{\psi}_\uparrow(\rr)
\hat{\psi}_\downarrow(\rr')+\text{H.c.}\right]
\nonumber\\ &\qquad
+ \frac{1}{2} \sum_{\sigma\sigma'} \mdint{3}{r}{r'} \hat{\psi}^\dagger_\sigma(\rr)\hat{\psi}^\dagger_{\sigma'}(\rr')
\frac{1}{|\rr-\rr'|} \hat{\psi}_{\sigma'} (\rr')\hat{\psi}_{\sigma}(\rr)
\nonumber\\ &\qquad
-\mdint{3}{r_1}{r_1'}\!\!\!\mdint{3}{r_2}{r_2'}\hat{\psi}_\downarrow^\dagger(\rr_1')\hat{\psi}_\uparrow^\dagger(\rr_1)
u(\rr_1',\rr_1,\rr_2,\rr_2')\hat{\psi}_\uparrow(\rr_2)
\hat{\psi}_\downarrow(\rr_2') \nonumber
\\ &\equiv \hat T  + \hat v - (\hat \Delta + \text{H.c.}) + \hat W  + \hat U.
\label{eq.hamiltonian}
\end{align}
\end{widetext}
The terms $\hat \Delta$ and $\hat \Delta^\dagger$ have no analog in
the non-superconducting electron gas.  As in BCS theory, a non-zero
expectation value for $\langle\hat\Delta\rangle$ must be understood as
a consequence of an external source field which is included by hand to
break the symmetry. It is switched off after the thermodynamic limit
is taken. In Eq.~\eqref{eq.hamiltonian}, $\hat T$ denotes the kinetic
energy operator, $\hat v$ the external potential, $\hat W$ the Coulomb
repulsion, while the operator $\hat U$ is an effective
electron-electron phonon-mediated attraction.

We consider systems described by such a Hamiltonian in a
grand-canonical ensemble at finite temperature $T=1/(k_\text{B}
\beta)$, where $k_\text{B}$ is the Boltzmann constant. The system is
coupled to a particle and a heat bath through the Lagrange multipliers
$\mu$ and $\beta$ (which we keep constant across this article). This
results in the well known formula for the grand potential:
\begin{align}
\Omega_{v,\Delta}\left[\varrho\right] = {\Tr}\,\hat{\varrho}\left(\hat{H}_{v,\Delta}-\mu \hat{N}
+\frac{1}{\beta}\ln \hat{\varrho}\right).
\end{align}
The reader should keep in mind that $\Omega_{v,\Delta}\left[\varrho\right]$
is dressed in the symbols $\beta$ and $\mu$. We refrain from writing them in order to alleviate the 
notation.
The minimizing equilibrium statistical density operator can be denoted as:
\begin{gather}
\hat{\varrho}_{\rm eq} = \frac1{\mathcal Z} e^{-\beta\left(\hat{H}_{v,\Delta}-\mu
\hat{N}\right)},
\label{eq:sdo}
\end{gather}
where ${\mathcal Z} = {\Tr}\big[e^{-\beta\left(\hat{H}_{v,\Delta}-\mu
\hat{N}\right)}\big]$.
Our objective is to prove (for fixed $\beta$ and $\mu$) a one-to-one 
relationship between the equilibrium statistical density operator 
$\varrho_{\rm eq}$, the external potentials $( v, \Delta)$, and the 
equilibrium Nambu-Gorkov one-body reduced density matrix $\Gamma_{\rm eq}$ 
or, equivalently, the pair $(\gamma_{\rm eq},\chi_{\rm eq})$. 
This mapping allows us to write all equilibrium observables as a functional 
of the equilibrium Nambu-Gorkov one-body reduced density matrix $\Gamma_{\rm eq}$. 

Before beginning the proof, we would like to comment on the problem of the phonons. 
In order to introduce consistently the electron-phonon interaction in SC-RDMFT we can 
follow the same recipe as in Ref.~\cite{SCDFT1,SCDFT2}. This involves treating the 
electronic and nuclear degrees of freedom at the same level, by introducing the
diagonal of the $N$-particle nuclear density-matrix as an extra variable, and by 
developing a multi-component RDMFT~\cite{PhysRevLett.86.2984}. The derivation is 
straightforward, and follows closely the one for SC-DFT, however complicates 
considerably the formulas. We therefore decided to overlook the nuclei, and only 
to present, in Sect.~\ref{decouplingP}, the electron-phonon term that stems from 
that formulation.

As a first step we prove (for fixed $\beta$ and $\mu$) the one-to-one
mapping between the equilibrium statistical density operator and the
pair of external potentials, namely:
\begin{equation}
{\varrho}_{\rm eq} \overset{1-1}{\longleftrightarrow} \left(v,
 \Delta\right).
\label{eq:1to1}
\end{equation}
Thus, the density $\varrho_{\rm eq}$ is uniquely determined by the potentials
$v$ and $\Delta$.
The proof goes by contradiction and is adapted from the original proof
for temperature-dependent RDMFT~\cite{PhysRevA.92.052514}.

We consider two Hamiltonians $\hat H_{v,\Delta}$ and $\hat H_{v',\Delta'}$ 
with the potentials $(v,\Delta)$ and $(v',\Delta')$ differing for more than a 
constant and assume that they lead to the same equilibrium statistical density operator. 
The following condition then holds:
\begin{gather}\label{conditiononeone}
\frac{e^{-\beta(\hat{H}_{v,\Delta}-\mu \hat{N})}}{\mathcal Z}=
\frac{e^{-\beta(\hat{H}_{v',\Delta'}- \mu \hat{N})}}{\mathcal Z'}.
\end{gather}
Here $\mathcal{Z}$ and $\mathcal{Z}'$ are the partition 
functions corresponding to the Hamiltonians $\hat H_{v,\Delta}$ and 
$\hat H_{v',\Delta'}$ respectively. Solving Eq.~\eqref{conditiononeone} 
for the difference of the Hamiltonians leads to:
\begin{align*}
\hat{H}_{v,\Delta}-\hat{H}_{v',\Delta'} = \hat v - \hat v' - (\hat \Delta - \hat \Delta '
+ {\rm h.c.})= \frac{1}{\beta}\ln\frac{\mathcal Z'}{\mathcal Z}.
\end{align*}
Since the right hand of this equation is particle conserving, 
we obtain trivially that $\hat\Delta = \hat\Delta'$. On the other hand,
$\hat v$ and $ \hat v'$ differ by a constant, which contradicts our initial 
assumption. Therefore, we obtain the desired 
one-to-one relationship~\eqref{eq:1to1}.

The second part of the theorem will follow the logic of the classic 
Mermin's proof for finite-temperature DFT~\cite{PhysRev.137.A1441}. 
It will yield (again) the one-to-one correspondence between the equilibrium
density operator $\varrho_{\rm eq}$ and $(\gamma_{\rm eq},\chi_{\rm eq})$. 
The proof goes (again) 
by \textit{reductio ad absurdum} as follows. Suppose that there exist two equilibrium 
density operators $ \varrho$ and $\varrho'$ that amount to the same 
$ \gamma$ and $\chi$. 
The Hamiltonian, statistical density operator, and grand potential associated 
with $v'$ and $\Delta'$ are labeled with $H_{v',\Delta'}$, $\varrho'$ 
and $\Omega_{v',\Delta'}$. It follows that:
\begin{align}
\label{eq:74}
\Omega_{v',\Delta'}\left[\varrho'\right] &= 
{\Tr}\hat \varrho'\left(\hat{H}_{v',\Delta'}-\mu \hat{N}
+\frac1\beta\ln \hat \varrho'\right) \\ &< \nonumber
{\Tr} \hat \varrho\left(\hat{H}_{v',\Delta'}-\mu \hat{N}
+\frac1\beta\ln \hat \varrho\right) \\ & = \nonumber
\Omega_{v,\Delta}\left[\varrho\right] 
+ {\Tr}[\hat{\varrho}(\hat v' - \hat v ) - \hat{\varrho}(\hat \Delta' - \hat \Delta + {\rm h.c.})]
\\ & = 
\Omega_{v,\Delta}\left[\varrho\right] 
+ v'[\gamma] - v[\gamma] - \Delta'\left[\chi\right] +
\Delta\left[\chi\right],
\nonumber
\end{align}
where
\begin{align*} 
v[\gamma] &\equiv \mdint{3}{r}{r'} v(\rr,\rr') \gamma(\rr,\rr'), \\
\Delta\left[\chi\right] &\equiv
\mdint{3}{r}{r'} \left[\Delta^*(\rr,\rr')\chi(\rr,\rr')+\Delta(\rr,\rr')\chi^*(\rr,\rr')\right].
\end{align*} 
Interchanging 
the primed and unprimed variables yields the following equation:
\begin{gather}
\Omega_{v,\Delta}\left[\varrho\right]<\Omega_{v',\Delta'}\left[\varrho'\right] -
 v'[\gamma] + v[\gamma] +
 \Delta'\left[\chi\right] - \Delta\left[\chi\right].
\label{eq:75}
\end{gather}
In \eqref{eq:75} we have used the hypothesis that both $\varrho$ 
and $\varrho'$ lead to the same reduced density matrices $\gamma$ 
and $ \chi$.
Adding Eq.~\eqref{eq:74} and \eqref{eq:75} one obtains a contradiction 
\textit{\`a la} Ho\-gen\-berg-Kohn, namely:
\begin{gather}
\Omega_{v,\Delta}\left[\varrho\right]+\Omega_{v',\Delta'}\left[\varrho'\right]
<\Omega_{v',\Delta'}\left[\varrho'\right]+\Omega_{v,\Delta}\left[\varrho\right],
\end{gather}
which is impossible. Therefore, the equilibrium redu\-ced densities 
$(\gamma_{\rm eq},\chi_{\rm eq})$  are unique for every 
equilibrium statistical density operator $\varrho_{\rm eq}$
(for fixed $\beta$ and $\mu$). Hence, there exists 
$\varrho_{\rm eq}[\gamma_{\rm eq},\chi_{\rm eq}]$,
a universal functional of the reduced densities.

Analogously to finite-temperature RDMFT we can write the potential
$\Omega_{v,\Delta}\left[\varrho_{\rm eq}\right]$ as a functional of 
$(\gamma_{\rm eq},\chi_{\rm eq})$, namely,
\begin{align}
\Omega^{\rm eq}_{v,\Delta} & \equiv \Omega[\varrho_{\rm eq}
[\gamma_{\rm eq},\chi_{\rm eq}]] \\
&= {\Tr} [(\hat T + \hat{v}-\mu)  \hat \gamma_{\rm eq}] - \Delta[\chi_{\rm eq}]
+\mathcal{F}_\beta\left[\gamma_{\rm eq},\chi_{\rm eq}\right].
\nonumber
\end{align}
Here we define the universal functional
$\mathcal{F}_\beta\left[\gamma_{\rm eq},\chi_{\rm eq}\right]$
whose existence is guaranteed by our two theorems. It reads
\begin{align}
\mathcal{F}_\beta\left[\gamma_{\rm eq},\chi_{\rm eq}\right] 
&={\Tr} \hat{\varrho}_{\rm eq}\left[\gamma_{\rm eq},\chi_{\rm eq}\right]\\
& \qquad\times
\left(\hat{W}+\hat{U} + \frac{1}{\beta}\ln \hat{\varrho}_{\rm eq}
\left[\gamma_{\rm eq},\chi_{\rm eq}\right]\right),\nonumber
\end{align}
where $\hat \varrho_{\rm eq}[\gamma_{\rm eq},\chi_{\rm eq}]$
is also a universal functional of the reduced densities. We should keep in mind 
that the functional $\mathcal{F}_\beta$ depends on $\beta$. Notice that, 
due to Eq.~\eqref{eq:Gamma}, the functional 
$\Omega_{v,\Delta}\left[\gamma_{\rm eq},\chi_{\rm eq}\right]$ can be written alternatively 
as a functional of the generalized one-body reduced density matrix 
(at equilibrium): $\Omega_{v,\Delta}\left[\Gamma_{\rm eq}\right]$.

Notice that up to this point our results are only defined on $\mathfrak{G}^{v,\Delta}$,
the domain of equilibrium $(v,\Delta)$-representable Nambu-Gorkov 
one-body reduced density matrices, defined as:
\begin{equation}
\mathfrak{G}^{v,\Delta}=\left\{\Gamma \in \mathfrak{G}^{N}
\, |\,\exists \varrho_{\rm eq} \: \text{such that} \:
\Gamma={\Tr}[ \hat \varrho_{eq} \hat \Gamma ]\right\},
\label{eqGamma}
\end{equation} 
with $\hat \Gamma =  \bar{\Psi}_{\sigma}(\rr) \otimes
        \bar{\Psi}^\dagger_{\sigma}(\rr').$
While this restriction is unproblematic for now, it will be a
cause of concern once we start any actual numerical minimization. 
The issue lies in the fact that  we are not able to characterize
in general the elements of $\mathfrak{G}^{v,\Delta}$. In order to 
overcome this challenge we will expand the minimization process to all 
ensemble-$N$-representable one-body reduced density matrices $\mathfrak{G}^{N}$,
namely, the set of generalized one-body reduced density matrices which are 
produced by a physical statistical density operator $\varrho\in \mathfrak{D}^N$ 
where 
\begin{equation}
    \mathfrak{D}^N = \{ \varrho\, | \,0 \leq \varrho \leq 1,\, {\Tr} \hat \varrho = 1,\,
{\Tr}[ \hat \varrho \hat N ] = N \}.
\end{equation}

In a fashion similar to Levy-Lieb's formulation of DFT \cite{Levy6062,doi:10.1002/qua.560240302}, 
we can restate the functional as a variational principle. Indeed, since by definition
\begin{align}
\Omega^{\rm eq}_{v,\Delta} = \min_{\varrho \in \mathfrak{D}^N}
\Omega_{v,\Delta}[\varrho],
\end{align}
we can divide the minimization as follows:
\begin{align}
\Omega^{\rm eq}_{v,\Delta} = \min_{\Gamma\in \mathfrak{G}^N} 
\inf_{\varrho \in \mathfrak{D}^N \rightarrow \Gamma}
\Omega_{v,\Delta}[\varrho].
\label{eq:func}
\end{align}
Now the functional is minimized on the set $\mathfrak{G}^N$, 
which can easily be characterized by the condition 
$0 \leq \Gamma\leq 1$~\cite{Bach1994,doi:10.1063/1.4941723}.

\subsection{Properties of the universal functional}

The infimum in Eq.~\eqref{eq:func} is justified by the fact that 
for a Nambu-Gorkov one-body reduced density matrix $\Gamma \in \mathfrak{G}^N$ 
that is non-$(v,\Delta)$-representable (i.e., it comes from a non-equilibrium 
statistical density operator), it is not clear if there exists a minimizing 
$ \varrho \rightarrow \Gamma$. We therefore prove the important theorem:
\begin{thm} 
For all $(\gamma,\chi)\in  \mathfrak{G}^N$, there exists 
$\varrho \in \mathfrak{D}^N$, such that $\varrho\rightarrow 
(\gamma,\chi)$ by contraction, and 
\label{inftomin2}
$$
\mathcal{F}_\beta\left[\gamma,\chi\right]=
{\Tr}\hat{\varrho}\left(\hat{W}+\hat{U}+\frac{1}{\beta}\ln\hat{\varrho}\right).
$$
\end{thm}

For the proof we require a sequence $\left\{\varrho_k\right\}$ 
with $\varrho_k\in \mathfrak{D}^N$ and $\varrho_k \rightarrow (\gamma,\chi)$.
Just as in Ref.~\cite{PhysRevA.92.052514} the proof is divided in three parts:

I.~\textit{$\exists\,\varrho : \varrho_k \rightarrow \varrho $ on the weak-*
topology.}

The proof of the first step remains the same as in 
Ref.~\cite{PhysRevA.92.052514} and therefore we refrain 
from repeating it. It has to be noted that it yields 
the equation:
\begin{equation}\label{eq:82}
{\Tr}[(\hat{\varrho}_k-\hat{\varrho})\hat{A}]\rightarrow 0
\end{equation}
for every compact operator $\hat{A}$. $ $\newline

II.~\textit{$\varrho$ yields the pair $(\gamma,\chi)$.}

For clarity's sake the reduced density matrices 
corresponding to $\varrho_k$ are labeled $\gamma_k$ and $\chi_k$ 
and the ones corresponding to $\varrho$ are denoted as 
$\bar{\gamma}$ and $\bar{\chi}$ even though they are all equal.
We will use the weak-* convergence of $\varrho_k$ to prove the 
weak-* convergence of $(\gamma_k,\chi_k)$ which yields the strong 
convergence. $\left\{\gamma_k\right\}$/$\left\{\chi_k\right\}$ 
is weakly-* convergent if and only if all linear functionals 
$\mathcal{L}\left[\gamma_k\right]$/$\mathcal{L}\left[\chi_k\right]$ 
on $\mathfrak{G}^N$ converge to $\mathcal{L}\left[\bar{\gamma}\right]$/
$\mathcal{L}\left[\bar{\chi}\right]$. In other words if for all bounded 
functions $f$:
\begin{equation}\label{eq:83}
  \abs{\mdint{3}{r}{r'} f(\rr,\rr')\left[\gamma_k(\rr,\rr')-\bar{\gamma}(\rr,\rr')\right]}\rightarrow 0
\end{equation}
and
\begin{equation}
  \label{eq:83b}
  \abs{\mdint{3}{r}{r'} f(\rr,\rr')\left[\chi_k(\rr,\rr')-\bar{\chi}(\rr,\rr')\right]+\text{H.c}}\rightarrow 0.
\end{equation}
In order to connect the anomalous densities with their statistical 
density operator we use the following operator:
\begin{gather}
\hat{O}_f=\mdint{3}{r}{r'} f(\rr,\rr')\hat\psi(\rr)\hat\psi(\rr')+\text{H.c.}
\end{gather}
Now, we can write Eq.~\eqref{eq:83b} as:
\begin{equation}\label{eq:86}
\abs{{\Tr}\big[(\hat{\varrho}_k-\hat{\varrho})\hat{O}_f\big]}\rightarrow 0.
\end{equation}
Unfortunately, we cannot apply Eq.~\eqref{eq:82} as $\hat{O}_f$ is 
not a compact operator. Nonetheless, as the eigenvalues of $\hat{O}_f$ are 
finite, the product of $\hat{O}_f$ with a finite dimensional projection 
operator $\hat{P}$ is compact. Using a projection $\hat{P}$ we transform 
the left hand of Eq.~\eqref{eq:86} to:
\begin{gather}
\label{eq:87}
\abs{{\Tr}\left[(\hat{\varrho}_k-\hat{\varrho})\hat{O}_f\hat{P}\right]+
{\Tr}\left[(\hat{\varrho}_k-\hat{\varrho})\hat{O}_f(1-\hat{P})\right]}\\
\leq\abs{{\Tr}\left[(\hat{\varrho}_k-\hat{\varrho})\hat{O}_f\hat{P}\right]}\nonumber
+\abs{{\Tr}\left[(\hat{\varrho}_k-\hat{\varrho})\hat{O}_f(1-\hat{P})\right]}.
\end{gather}
We can show separately that each part goes to zero. By 
applying Eq.~\eqref{eq:82} to the left part of Eq.~\eqref{eq:87} 
we can see that it goes to zero. For the right part we prove separately 
that the terms with $\varrho$ and $\varrho_k$ both go to zero. 
To start with, we can choose $\hat P$ such that:
\begin{gather}
\overset{\infty}{\underset{i=M+1}{\sum}} w_{ki}\bra{\Psi_{ki}}\hat{O}_f\ket{\Psi_{ki}}<\epsilon.
\end{gather}
where M depends on the dimension of $\hat{P}$ and $w_{ki}$ are the 
eigenvalues of $\varrho_k$. This means we can choose a $K$ such 
that 
$$
\abs{{\Tr}\left[(\hat{\varrho}_k-\hat{\varrho})\hat{O}_f\hat{P}\right]}
< \epsilon
$$ for all $k\geq K$ and $\epsilon \geq 0$ and we
 can choose an $M$ such that the same is true for 
$$
\abs{{\Tr}\left[(\hat{\varrho}_k-\hat{\varrho})\hat{O}_f(1-\hat{P})\right]}.
$$ 
The same process can be repeated for $\gamma$. Indeed, by choosing the larger 
$K$ and $M$ from the two proofs we fulfill the conditions for the weak 
convergence of the pair $(\gamma_k,\chi_k)$.
In other words the weak-* convergence of $\varrho_k$ implies the weak convergence 
of $(\gamma_k,\chi_k)$ and because we choose all $(\gamma_k,\chi_k)=(\gamma,\chi)$ 
it implies strong convergence and $\varrho\rightarrow (\gamma,\chi)$.

III.~\textit{ $\mathcal{F}_\beta[\gamma,\chi] = {\Tr}\hat \varrho\left(
\hat{H}+\frac{1}{\beta}\ln\hat{\varrho}\right)$ for this $\varrho$.}

The third and last step is once again completely equivalent to \cite{baldsiefen2012reduced}.

Therefore, there is in fact a minimizing $\varrho$ in Eq.~\eqref{eq:func} for all
$\Gamma \in \mathfrak{G}^N$, and we can replace the infimum by a minimum,
and the functional can be written as
\begin{align}
\Omega^{\rm eq}_{v,\Delta} = \min_{\Gamma\in \mathfrak{G}^N} 
\min_{\varrho \in \mathfrak{D}^N \rightarrow \Gamma}
\Omega_{v,\Delta}[ \varrho].
\label{eq:func2}
\end{align}
This extension allows us to define a Levy-Lieb functional for SC-RDMFT, 
namely:
\begin{align}
\label{eq:Lieb}
\mathfrak{F}_\beta[\gamma,\chi] =
\underset{\varrho\rightarrow (\gamma,\chi)}{\min}
{\Tr} \hat{\varrho}\left(\hat{W}+\hat U + \frac{1}{\beta}\ln\hat{\varrho}\right).
\end{align}
This functional justifies functional differentiability within
SC-RDMFT, which is required both to develop a Kohn-Sham framework and
to engineer exchange-correlation functionals.  Since
$\Gamma$ and $(\gamma,\chi)$ are uniquely determined we will not
distinguish between $\mathfrak{F}_\beta[\gamma,\chi]$ and
$\mathfrak{F}_\beta[\Gamma]$ (or from $\mathcal{F}_\beta[\gamma,\chi]$ and
$\mathcal{F}_\beta[\Gamma]$), from now on.  A second important property
of the functional \eqref{eq:Lieb} is its convexity, which follows
directly from the convexity of the grand potential
$\Omega_{v,\Delta}[\gamma,\chi]$.

A crucial problem in reduced density theories (DFT or RDMFT) 
is the problem of $v$-representability. We already defined the set 
$\mathfrak{G}^{v,\Delta} \subset \mathfrak{G}^N$ in Eq.~\eqref{eqGamma} 
as the set of $(v,\Delta)$-representable Nambu-Gorkov reduced density 
matrices. For the following we will use a different but equivalent definition:
$(\gamma_0,\chi_0) \in \mathfrak{G}^{v,\Delta}$ if there exists 
$(v,\Delta)$ such that
\begin{align}
\nonumber
& \underset{(\gamma,\chi) \in \mathfrak{G}^N}{\inf}\left(\mathfrak{F}_\beta\left[
\gamma,\chi\right]
+ {\Tr}[(\hat T + \hat v -\mu) \hat\gamma] - \Delta[\chi] \right)
\\ &\qquad =  \mathfrak{F}_\beta\left[\gamma_0,\chi_0\right] + 
{\Tr}[(\hat T + \hat v - \mu) \hat\gamma_0] -  \Delta[\chi_0].
\end{align}

Remarkably, $\mathfrak{G}^{v,\Delta}$ is dense in $\mathfrak{G}^N$.
This result is crucial when deriving functionals through many-body
perturbation theory, as we will do in the following. For instance, in
$\mathfrak{G}^{v, \Delta}$ we can relatively safely assume that a
grand potential approximation using a functional derived through
many-body perturbation theory, specifically the Sham-Schl\"uter
connection \cite{ss1,ss2,ss3}, is minimized by the correct equilibrium
Nambu-Gorkov one-body density matrix. We already discussed that, in
practice, a minimization has to be carried out in the set of
ensemble-$N$-representable Nambu-Gorkov one-body reduced density
matrices $\mathfrak{G}^N$.
 
For the proof of the density of $\mathfrak{G}^{v,\Delta}$ on 
$\mathfrak{G}^{N}$
we require the concept of a continuous tangent functional (CTF):
Consider a real functional $\mathcal{F}$ on a subset $\mathcal{S}$ of a Banach 
space $\mathfrak{H}$. Let $x_0\in \mathcal{S}$. A continuous linear func\-tional 
$\mathcal{L}$ is said to exhibit a CTF at $x_0$ if \cite{Engel}:
\begin{gather}
\forall x \in \mathcal{S}:\, \mathcal{F}\left[x\right]\geq 
\mathcal{F}\left[x_0\right]-\mathcal{L}\left[x-x_0\right].
\end{gather}
The crucial observation is the following theorem:
\begin{thm}
\label{thm2}
The universal functional $\mathfrak{F}_\beta$ 
has a unique CTF at every pair $(\gamma,\chi)\in \mathfrak{G}^{v,\Delta}$ 
and only at these pairs. 
\end{thm}
The proof of this latter statement follows 
by \textit{reductio ad absurdum}. 
Consider a pair $(\gamma',\chi')$ that is not $(v,\Delta)$-representable 
and suppose that $\mathfrak{F}_\beta\left[\gamma',\chi'\right]$ has a CTF denoted as 
$(\tilde{v},\tilde{\Delta})$. Therefore,
\begin{align}
\mathfrak{F}_\beta\left[\gamma,\chi\right]\geq 
\mathfrak{F}_\beta\left[\gamma',\chi'\right]-
\tilde{v}[\gamma-\gamma']+\tilde{\Delta}[\chi-\chi'].
\end{align}
Notice the change of sign in front of the anomalous term in
the Hamiltonian \eqref{eq.hamiltonian}.
Thus,
\begin{align}
&\underset{(\chi,\gamma)\in \mathfrak{G}^N}{\inf} \nonumber
\left(\mathfrak{F}_\beta\left[\gamma,\chi\right]+ 
\tilde{v}[\gamma]- \tilde{\Delta}[\chi]\right) \\ & \qquad \qquad \geq 
\mathfrak{F}_\beta\left[\gamma',\chi'\right]+ \tilde{v}[\gamma']- \tilde{\Delta}[\chi'].
\end{align}
The infimum is only assumed for $(\chi,\gamma)\in \mathfrak{G}^{v,\Delta}$
and only then is the equality fulfilled. This contradicts our assumption 
that $(\gamma',\chi')$ is not equilibrium $(v,\Delta)$-representable.
In this sense, the existence of a CTF at $(\gamma',\chi')$ implies the 
equilibrium $(v,\Delta)$-representability of $(\gamma',\chi')$.
The opposite direction follows trivially from the definition of 
equilibrium $(v,\Delta)$-representability.

One last theorem, a less known variant of the famous 
Bishop-Phelps theorem~\cite{bishop1986proof}, is required 
to conclude the proof.
Let $\mathcal{F}$ be a lower semi-continuous convex functional 
on a real Banach space $\mathfrak{H}$. Suppose $x_0=(\gamma_0,\chi_0)
\in \mathfrak{H}$ and $V_0=(v_0,\Delta_0)\in \mathfrak{H}^*$. 
Then, for every $\epsilon > 0$, there exists $x_{\epsilon} \in 
\mathfrak{H}$ and $V_{\epsilon} \in \mathfrak{H}^*$ such that~\cite{bishop1986proof}:
\begin{enumerate}
\item $||V_{\epsilon}-V_{0}|| \leq \epsilon$.
\item $V_{\epsilon}$ is CTF to $\mathcal{F}$ at $x_\epsilon$.
\item $\epsilon ||x_{\epsilon}-x_{0}|| \leq \mathcal{F}\left[x_0\right]
+ v_0[\gamma_0] + \Delta_0[\chi_0] - \underset
{x \in \mathfrak{H}}{\inf}\left(\mathcal{F}\left[x\right]+
v_0[\gamma]+\Delta_0[\chi]\right)$.
\end{enumerate}

This theorem can be used to prove the following result, 
which is equivalent to our hypothesis that $\mathfrak{G}^{v,\Delta}$ 
is dense in $\mathfrak{G}^N$:

\begin{thm}
For all $x\in \mathfrak{G}^N$ there exists a sequence $\left\{x_k\right\}\subset 
\mathfrak{G}^N$, such that
$x_k\rightarrow x$ and $x_k\in \mathfrak{G}^{v,\Delta}, \,\forall\,k$.
\end{thm}

To prove this, we start by denoting the right side of the third condition 
in the Bishop-Phelps Theorem as $C_0$, which is independent of $\epsilon$ and finite. 
Let $\epsilon_k = k C_0$. We can find an $x_k$ for each $\epsilon_k$ at which 
$\mathcal{F}$ has a CTF such that 
\begin{equation}
||x_{k}-x_0|| \leq \frac{C_0}{\epsilon_k}=\frac{1}{k}.
\end{equation}
We know from Theorem \ref{thm2} that the existence of a CTF 
to $\mathcal{F}$ at $x_{k}$ is equivalent to the equilibrium 
$(v,\Delta)$-representability of $x_k$. This concludes not 
only the proof of the theorem but also our original task of proving 
that the set $\mathfrak{G}^{v,\Delta}$ is dense in the set $\mathfrak{G}^N$.

\section{Kohn-Sham system for SC-RDMFT}
\label{KohnSham}

After deriving a Hohenberg-Kohn theorem, the next logical step is to
discuss a Kohn-Sham system for SC-RDMFT. We develop two different
approaches to accomplish such a goal. The first is analogous to SC-DFT
(or normal DFT as well) and works through functional
derivatives. However, just like in SC-DFT it does not prove the
existence of the Kohn-Sham system. The second approach, on the other
hand, is analogous to finite-temperature
RDMFT and yields a proof for the existence of
the Kohn-Sham system.

\subsection{Approach 1: analogous to SC-DFT}

The following discussion is only valid at finite temperature, 
since no Kohn-Sham system exists at zero temperature \cite{firstpaper}.
In SC-DFT a Kohn-Sham system is defined as a non-interacting system that 
reproduces the electronic density $n_{\rm eq}$ and the anomalous density 
$\chi_{\rm eq}$ of
the interacting system. In the same vein, we define the Kohn-Sham system
 in SC-RDMFT as the non-interacting system with the same (equilibrium) 
reduced density matrices $(\gamma, \chi)$. For systems with spin-rotational 
symmetry, the Nambu-Gorkov one-body reduced density matrix
\[
\Gamma(\rr,\rr') =\begin{bmatrix}
\gamma(\rr,\rr') & \chi^{\dagger}(\rr,\rr') \\ \chi(\rr,\rr') & \delta(\rr-\rr')-\gamma(\rr',\rr)
\end{bmatrix}
\]
is equivalent to such densities. As a result of the one-to-one relation
\eqref{eq:1to1} the grand potential can be written as: 
\begin{gather}
\Omega\left[\gamma,\chi\right]=
\mathfrak{F}_\beta\left[\gamma,\chi\right] +
{\Tr} [(\hat T + \hat{v}-\mu)  \hat \gamma] - 
\Delta[\chi],
\end{gather}
where 
\begin{equation}
\Delta[\chi] \equiv
\mdint{3}{r}{r'} \left[\Delta^*(\rr,\rr')\chi(\rr,\rr')+\Delta(\rr,\rr')\chi^*(\rr,\rr')\right].
\end{equation}
We divide the universal functional in meaningful parts:
\begin{align}
\mathfrak{F}_\beta\left[\gamma,\chi\right]
=-\frac{1}{\beta}S_{\text{KS}}\left[\gamma,\chi\right]+
U_{\text{Hartree}}\left[\gamma\right]+
\mathcal{F}_{\text{\text{xc}}}\left[\gamma,\chi\right],
\nonumber
\end{align}
where $U_{\text{Hartree}}$ is the Hartree term of the 
total electron-electron interaction energy and $S_{\text{KS}}$ 
is the entropy of the Kohn-Sham system. A natural advantage 
of RMDFT, unlike DFT, is of course our explicit knowledge 
of the kinetic energy functional. 

We denote the grand-canonical potential of the Kohn-Sham system as:
\begin{equation}
\Omega_{\text{KS}}\left[\gamma,\chi\right]= 
{\Tr} [(\hat T + \hat{v}_{\rm KS}-\mu)  \hat \gamma] - 
\Delta_{\rm KS}[\chi] -\frac{1}{\beta}S_{\rm KS}\left[\gamma,\chi\right].
\end{equation}
Our aim is now to determine the Kohn-Sham potentials $v_{\text{KS}}$ 
and $\Delta_{\text{KS}}$ such that the Kohn-Sham grand canonical 
potential $\Omega_{\text{KS}}\left[\gamma,\chi\right]$ is minimized 
by the same $(\gamma,\chi)$ of the interacting system.
As in normal DFT, at this point one needs the assumption that at 
finite temperatures the two relevant interacting densities $\gamma$ 
and $\chi$ are smooth  (i.e.,  normalizable  and  twice differentiable)
and therefore noninteracting $(v,\Delta$)-representable \cite{Engel}.
We then obtain:
\begin{multline}
\frac{\delta \Omega\left[\gamma,\chi\right]}{\delta \gamma(\rr,\rr')} 
=\frac{\delta \mathcal{F}_{\text{xc}}\left[\gamma,\chi\right]}{\delta \gamma(\rr,\rr')}
+[v(\rr,\rr') -\mu\delta(\rr-\rr')]
\\ +\mint{3}{r'} \frac{\gamma(\rr,\rr)}{\abs{\rr-\rr'}}
+\frac{\delta T\left[\gamma\right]}{\delta \gamma(\rr,\rr')}
-\frac{1}{\beta}\frac{\delta S_{\text{KS}}[\gamma,\chi]}{\delta \gamma(\rr,\rr')}
\end{multline}
and 
\begin{multline}
\frac{\delta \Omega_{\text{KS}}\left[\gamma,\chi\right]}{\delta \gamma(\rr,\rr')}
= [v_{\text{KS}}(\rr,\rr')-\mu \delta(\rr-\rr')]
\\
+\frac{\delta T\left[\gamma\right]}{\delta \gamma(\rr,\rr')}
-\frac{1}{\beta}\frac{\delta S_{\text{KS}}[\gamma,\chi]}
{ \delta \gamma(\rr,\rr')}
\end{multline}
and therefore
\begin{align}
v_{\text{KS}}(\rr,\rr') = v(\rr,\rr')+\mint{3}{r'} \frac{\gamma(\rr,\rr)}{\abs{\rr-\rr'}}+v_{\text{xc}}(\rr,\rr'),
\end{align}
with $v_{\text{xc}}(\rr,\rr') = \delta \mathcal{F}_{\text{xc}}/\delta \gamma(\rr,\rr')$.
The Kohn-Sham potential consists of an external potential, a Hartree term,
and the exchange-correlation term. After analogous calculations for 
$\chi(\rr,\rr')$ we arrive at:
\begin{gather}\label{anomalouskspot}
\Delta^*_{\text{KS}}(\rr,\rr')
= \Delta^*(\rr,\rr') -\frac{\delta \mathcal{F}_{\text{xc}}}{\delta \chi(\rr,\rr')}.
\end{gather}
Notice that the universal exchange-correlation functional energy 
$\mathcal{F}_{\text{xc}}$ is comprised of first the exchange-correlation 
energy of the electrons and second the difference between the interacting 
and non-interacting entropy. The existence of the Kohn-Sham potentials 
hinges on the existence of the functional derivatives of 
$\mathcal{F}_{\rm xc}\left[\gamma,\chi\right]$. Although 
we cannot answer this question in general, in practice we will use 
differentiable approximations for $\mathcal{F}_{\text{xc}}$.

Diagonalizing the Kohn-Sham Hamiltonian by a Bogoliubov
transformation of the field operators $\hat \psi_\sigma$:
\begin{align}
\label{eq:BdGtrnasf}
\hat\psi_{\sigma}(r) = 
\sum_{i} u_{i}(x) \hat\gamma_{i,\sigma}-\text{sign}(\sigma)v^*_{i}(x) 
\hat\gamma^{\dagger}_{i,-\sigma},
\end{align}
where $\hat\gamma^\dagger_{k,\sigma}$ and $\hat\gamma_{k,\sigma}$ 
are creation and annihilation operators of fermionic quasiparticles,
yields the following Bogoliubov-de-Gennes equations which are 
completely analogous to SC-DFT and only differ in the non-locality 
of the Kohn-Sham potential:
\begin{widetext}
\begin{align}\label{BGDE}
-\left[\frac{\grad^2}{2}+\mu\right]u_i(\rr) + \mint{3}{r'} v_{\text{KS}}(\rr,\rr') u_i(\rr')
+\mint{3}{r'} \Delta_{\text{KS}}(\rr,\rr')v_i(\rr') &=\epsilon_i u_i(\rr)\\
\nonumber\\
\left[\frac{\grad^2}{2}+\mu\right]v_i(\rr) - \mint{3}{r'} v_{\text{KS}}(\rr,\rr')v_i(\rr')
+ \mint{3}{r'} \Delta_{\text{KS}}^*(\rr,\rr')u_i(\rr') &=\epsilon_i v_i(\rr).
\end{align}

These are the self-consistent SC-RDMFT equations for the Kohn-Sham orbitals.
In \hyperref[AppendixA]{Appendix A} we present the full Nambu-Gorkov Green's functions
for SC-RDMFT.
\end{widetext}

\subsection{Approach 2: analogous to finite-temperature RDMFT}
\label{KSapproach2}

In finite temperature RDMFT the existence of a Kohn-Sham system is 
proven by introducing a common system of eigenfunctions of the 
Kohn-Sham Ha\-mil\-tonian and the one-body reduced density matrix. 
Unfortunately, the Ha\-mil\-tonian of the superconducting Kohn-Sham 
system contains two anomalous terms ($\Delta$ and $\Delta^*$)
which do not commute with the particle-number operator $\hat N$. 
However, we can derive the Kohn-Sham system through a second more 
original approach we shortly describe in this subsection, by directly 
searching for the Bogoliubov transform that diagonalizes $\Gamma$ in 
the following form:
\begin{equation}
\hat{W}^{\dagger}\Gamma\hat{W}=\Gamma_{\rm diag}=
\begin{bmatrix}\lambda_i&0\\0&1-\lambda_i\end{bmatrix}.
\end{equation}
Such a Bogoliubov transform exists for every ensemble $N$-representable 
$\Gamma$ \cite{Bach1994}. In this new auxiliary Bogoliubov-de-Gennes system, 
represented by only $\gamma$ (as the anomalous terms are zero) we can 
basically return to the normal finite-temperature RDMFT and the corresponding 
proof for the existence of a Kohn-Sham system~\cite{PhysRevA.92.052514}. We 
define new field operators belonging to the creation and annihilation 
operators $\hat{\gamma}_i$ and $\hat{\gamma}^\dagger_i$ of the Bogoliubov 
quasiparticles:
\begin{gather}
\tilde{\Psi}(\rr)=\sum_i \gamma_i \tilde{\phi}_i(\rr) \quad {\rm and}
\quad \tilde{\Psi}^\dagger(\rr)=\sum_i \gamma^\dagger_i \tilde{\phi}^*_i(\rr),
\end{gather}
where $\tilde{\phi}$ are the natural orbitals (eigenfunctions) of $\Gamma_{\rm diag}$. 
The density matrix $\gamma(\rr,\rr')$ is defined as:
\begin{align}
\gamma(\rr,\rr') &= \text{Tr}\left[\frac{e^{-\beta(\hat{H}-\mu\hat{N})}}{\mathcal{Z}}\hat{\tilde{\Psi}}^\dagger(\rr')\hat{\tilde{\Psi}}(\rr)\right] 
\\ &=\sum_i\frac{1}{1+e^{\beta(\epsilon_i-\mu)}}\tilde{\phi}^*_i(\rr')\tilde{\phi}_i(\rr)=\sum_i\lambda_i\tilde{\phi}^*_i(\rr')\tilde{\phi}_i(\rr). \nonumber
\end{align}
These equations highlight that the occupation numbers 
of the Bogoliubov quasiparticles $\lambda_i$ belong to eigenenergies 
$\epsilon_i$ and are connected through a Fermi-Dirac distribution:
$\lambda_i= 1 /(1+e^{\beta(\epsilon_i-\mu)})$.
Thus we arrive at an explicit equation for a Kohn-Sham potential 
in the Bogoliubov-de-Gennes system:
\begin{gather}
v^{B}_{\text{KS}}(\rr,\rr')=\sum_{i,j} \left(\delta_{ij}\epsilon_i-t_{ij}\right)\tilde{\phi}^*_i(\rr')\tilde{\phi}_j(\rr),
\end{gather}
where $t_{ij}$ are the entries of the kinetic-enegy operator. 
Obviously, the Hamiltonian can be written as 
$\hat{H}=\sum_i \epsilon_i \hat{\gamma}_i^\dagger\hat{\gamma}_i$.
We now recognize this form as the result of the Bogoliubov transform 
of the superconducting Kohn-Sham Hamiltonian. If we reverse the Bogoliubov 
transform that diagonalized $\Gamma$ we arrive at Kohn-Sham potentials that 
fulfill our familiar Bogoliubov-de-Gennes equations~\eqref{BGDE}.
Consequently, unlike SC-DFT, we can solve the problem of non-interacting 
$(v,\Delta)$-representability at finite temperature in SC-RDMFT as we can 
prove that a Kohn-Sham system exists for every $\Gamma\in \mathfrak{G}^N$ at finite 
temperature. 

Furthermore, due to the one-to-one mapping between 
$(v,\Delta)$, $(\gamma,\chi)$, and $\varrho_{\rm eq}$ we know that both ways of reaching a Kohn-Sham system result 
in exactly the same Kohn-Sham potentials and equilibrium state. It is worth 
noticing that at zero temperature the situation is completely different. 
Indeed, since at zero temperature the ground state of a normal 
non-interacting system is a Slater determinant, the corresponding one-body 
reduced density matrix is idempotent. The same holds for a superconducting 
system: it turns out that the Nambu-Gorkov one-body reduced density matrix 
is also idempotent, which implies:
\begin{gather}
\gamma=\gamma^2+\chi^\dagger \chi.
\end{gather}
Therefore, $\chi$ and $\gamma$ are not independent. However, 
in perturbation theory, for an interacting system $\Gamma$ is idempotent in 
first order and only ceases to be so in second order \cite{firstpaper}. 
The idempotence of $\Gamma$ in first-order 
perturbation theory hints at the fact that even though a Kohn-Sham system does 
not exist at zero temperature, a practical implementation might nevertheless 
not fail completely.

\section{Decoupling Approximation}
\label{decoupling}

In this section we develop a useful computational framework of SC-RDMFT.
As a starting point, we solve the electronic problem with the Kohn-Sham 
potential $v_{\text{KS}}\left[\gamma,\chi\right]$.
Assuming a crystal lattice we can write the natural orbitals 
as Bloch states.
Furthermore, we denote the functions $u_i(\rr)$, $v_i(\rr)$, and $\Delta(\rr,\rr')$ 
in the basis of the natural orbitals of the non-superconducting solution, namely:
\begin{gather}
u_i(\rr)=\sum_{\nk}u_{i;\nk}\phi_{\nk}(\rr)\\
v_i(\rr)=\sum_{\nk}v_{i;\nk}\phi_{\nk}(\rr)\\
\Delta_{\text{KS}}(\rr,\rr')=\sum_{\nk n'\bm{k}'}\Delta_{\text{KS};\nk n'\bm{k}'}\phi^*_{n'\bm{k}'}(\rr')\phi_{\nk}(\rr).
\end{gather}
In what follows, we will use this eigenbasis of natural Bloch orbitals and the solution of 
the normal electronic system to solve the phononic and superconducting problem.

\subsection{Electron-phonon coupling}
\label{decouplingP}

Several approximations are needed in order to treat phonons in SC-RDMFT. 
First of all, it is reasonable to assume that the atoms only move from 
their equilibrium lattice positions through small oscillations and therefore 
the usual harmonic approximation applies. Second, the lattice dynamics will 
be approximated by the ones of the corresponding non-superconducting system. In this way, we can define 
the electron-phonon scattering matrix elements in a completely analogous way to 
SC-DFT~\cite{sanna2}:
\begin{gather}
\label{pecooupl}
g^{\nu}_{m\bm{k}+\bm{q},\nk}=\sqrt[]{\frac{\hbar}{2\omega_{\bm{q}\nu}}}\bra{\phi_{m\bm{k}+\bm{q}}}
\Delta V_{\rm scf}^{\bm{q}\nu} \ket{\phi_{\nk}}
\end{gather} 
The main difference lies in the fact that $\phi_{\nk}$ are not DFT-Kohn-Sham 
orbitals but natural orbitals. In addition, the Kohn-Sham potential is non-local.
In Eq.~\eqref{pecooupl}, $\bm{q}$ and $\bm{k}$ are the phonon and electron momenta, 
$n$  and $m$
are the natural-orbital band indices, $\omega_{\nu}$ the phonon frequency, 
$\nu$ the phonon branch and $\Delta V_{\rm scf}^{\bm{q}\nu}$ the variation in the 
Kohn-Sham potential due to ionic displacement. This results in the Hamiltonian:
\begin{align}
 H_{\rm e-ph}  =\sum_{mn\sigma,\nu \bm{kq}}g^{\nu}_{m\bm{k}+\bm{q},\nk}
\hat\psi^\dagger_{\sigma m\bm{k}+\bm{q}}\psi_{\sigma \nk}
(\hat b_{\nu \bm{q}}+\hat b^\dagger_{\nu -\bm{q}}) ,
\end{align}
where $\hat b_{\nu \bm{q}}$ and $\hat b^\dagger_{\nu \bm{q}}$ are the
annihilation and creation operators of the phonons.  The form of the
electron-phonon coupling in SC-RDMFT and SC-DFT seems very
similar. However, we should notice that the electron-phonon coupling
constants that enter our theory should, in principle, be calculated
from RDMFT. While in DFT the matrix elements
$g^{\nu}_{m\bm{k}+\bm{q},\nk}$ are easily obtained from density
functional perturbation theory~\cite{phonons}, such a framework still
does not exist for RDMFT for solids. Yet, as  discussed  
in SC-DFT \cite{SCDFT1},  to solve the corresponding gap equation are 
required the  electron-phonon coupling constants, $g^{\nu}_{m\bm{k}+\bm{q},\nk}$, 
as well as the normal-state Kohn-Sham eigenenergies $\xi_{\nk}$. In practice, however,
the phononic contributions to the functionals are often averaged 
on the Fermi surface \cite{SCDFT2}:
\begin{align}
\sum_{m\nk } \sum_{\nu,\bm{q}} \left|g^{\nu}_{m\bm{k}+\bm{q},\nk}\right|^2
\delta(\xi_{nk})\delta(\xi_{m \bm{k}+\bm{q}}).
\end{align}

\subsection{Band-decoupling approximation}

In SC-DFT the decoupling approximation is based on the assumption 
that the superconducting transition does not introduce any structural 
transition. It is also assumed that it does not cause any hybridization 
between the bands. We already assumed that the first approximation is true 
when we treated the phonons. However, the first approximation also includes 
the fact that $\Delta_{\text{KS}}(\rr,\rr')$ has the periodicity of the lattice 
and it is therefore unnecessary to sum over the indices $k$. The second 
assum\-ption 
reduces the equations for $u_i(\rr)$ and $v_i(\rr)$ to:
\begin{align}
  u_i(\rr) & \equiv u_{\nk}(\rr) = u_{\nk}\phi_{\nk}(\rr)  \nonumber \\
  v_i(\rr) & \equiv v_{\nk}(\rr) = v_{\nk}\phi_{\nk}(\rr).
\end{align}
This approximation is motivated by the difference in energy scale
between the electronic bonding and superconducting pairing. If any of
the bands are degenerate on the scale of $\Delta_{\text{KS}}$ one can
question the validity of the approximation. Yet, even then, we expect
the hybridization to be negligible.

Inserting these approximations into the Bogoliubov-de-Gennes
equations~\eqref{BGDE}, and using the orthogonality of the basis set
we arrive at:
\begin{align}
\tilde{\epsilon}_{\nk}u_{\nk}+\Delta_{\KS;\nk} v_{\nk}&=E_{\nk}u_{\nk}\nonumber\\
-\tilde{\epsilon}_{\nk}v_{\nk}+\Delta^*_{\KS;\nk} u_{\nk}&=E_{\nk}v_{\nk}.
\end{align}
The energies $\tilde{\epsilon}_{\nk}$ are defined as the energies from 
the non-superconducting RDMFT calculation minus the chemical potential.
This yields equations of the same form as in SC-DFT. Indeed, for the 
eigenenergies we obtain:
\begin{align}
E_{\nk}=\pm\sqrt[]{\tilde{\epsilon}_{\nk}^2+\abs{\Delta_{\KS;\nk}}^2}
\end{align}
and for the amplitudes:
\begin{align}
u_{\nk}&=\frac{1}{\sqrt[]{2}}\text{sign}(E_{\nk})e^{\Phi_{\nk}}
\sqrt[]{1+\frac{\tilde{\epsilon}_{\nk}}{\abs{E_{\nk}}}}, \nonumber\\
v_{\nk}&=\frac{1}{\sqrt[]{2}}\sqrt[]{1-\frac{\tilde{\epsilon}_{\nk}}{\abs{E_{\nk}}}},
\end{align}
where $e^{\Phi_{\nk}}$ is the phase of $\Delta_{\KS;\nk}$. 
We are already familiar with the first equation from BCS-theory 
where $\Delta_{\KS,\nk}$ is the energy gap. We will interpret 
$\Delta_{\KS,\nk}$ in the same fashion in SC-RDMFT.
Of course the solution of the Bogoliubov-de-Gennes equations also 
results in new formulas for $\gamma$ and $\chi$. Here we use the fact 
that the Bogoliubov-quasiparticle states are occupied according to a Fermi-Dirac distribution.
The densities $\gamma$ and $\chi$ take on the form:
\begin{align}
\label{gap}
\gamma(\rr,\rr')&=\sum_{\nk} \abs{u_{\nk}}^2\frac{1}{e^{\beta E_{\nk}}+1}
\phi_{\nk}^*(\rr')\phi_{\nk}(\rr) \\&\quad +\abs{v_{\nk}}^2
\left(1-\frac{1}{e^{\beta E_{\nk}}+1}\right)\phi_{\nk}^*(\rr)\phi_{\nk}(\rr')
\nonumber\\
&=\sum_{\nk}\left[ \frac{1}{2}-\frac{\tilde{\epsilon}_{\nk}}{\abs{E_{\nk}}}\tanh\left(\frac{\beta \abs{E_{\nk}}}{2}\right)\right]\phi^*_{\nk}(\rr')\phi_{\nk}(\rr) \nonumber
\end{align}
and
\begin{align}
\chi(\rr,\rr')=\frac{1}{2}\sum_{\nk} \frac{\Delta_{\KS;\nk}}{\abs{E_{\nk}}}\tanh\left(\frac{\beta \abs{E_{\nk}}}{2}\right)\phi^*_{\nk}(\rr')\phi_{\nk}(\rr).
\end{align}
We defined the anomalous exchange-correlation potential 
$\Delta_{\text{xc}}$ as the functional derivative of the 
universal functional with respect to $\chi$. This, 
in combination with Eq.~\eqref{gap}, yields a self 
consistent equation for the anomalous exchange-correlation potential:
\begin{equation}
\Delta_{\text{xc}}=-\frac{\delta F_{\text{xc}}
\left[\gamma,\chi\left[\Delta_{\text{KS}},\gamma\right]\right]}{\delta \chi}.\label{eq:118}
\end{equation}
Here we used the fact that $\chi$ is a functional of $\Delta$
 and $\gamma$ and consequently we can calculate the functional 
derivative in Eq.~\eqref{eq:118} through the chain rule as a 
derivative with respect to $\Delta_{\text{KS}}$ and $\gamma$. 
As we consider $\Delta(\rr,\rr')$ as the superconducting energy gap 
we arrive at a self-consistent gap equation which we 
can solve in order to determine the transition temperature.

\section{The Sham-Schl\"uter connection}\label{SSC}

The last missing piece in the SC-RDMFT puzzle is an expression 
for the universal exchange-correlation functional. The Sham-Schl\"uter 
connection~\cite{ss1,ss2,ss3} is a many-body perturbation approach 
for the derivation of exchange-correlation potentials. The idea was 
already introduced to SC-DFT in the original paper by Oliveira, Gross 
and Kohn~\cite{PhysRevLett.60.2430}. We will use the existence of the 
Kohn-Sham system to show that the Sham-Schl\"uter connection can be easily 
extended to SC-RDMFT. 

The starting point is the Dyson equation in Nambu-Gorkov space:
\begin{gather}
\bar{G}=\bar{G}^{\text{KS}}+\bar{G}^{\text{KS}}\bar{\Sigma}^{\text{KS}}\bar{G},
\end{gather}
where $\bar{G}^{\text{KS}}$ is the non-interacting Green's function which 
corresponds to the SC-RDMFT Kohn-Sham Hamiltonian:
\begin{gather}
\bar{H}_{\text{KS}}(\rr,\rr')=
\begin{bmatrix}
h_{\text{KS}}(\rr,\rr') &\Delta_{\text{KS}}(\rr,\rr')\\
\Delta^*_{\text{KS}}(\rr,\rr')&-h_{\text{KS}}(\rr,\rr')
\end{bmatrix}.
\end{gather}
$\bar{\Sigma}^{\text{KS}}(\rr,\rr',\omega_i)$ is the self energy 
of the system:
\begin{equation}
\bar{\Sigma}^{\text{KS}}(\rr,\rr',\omega_i)=\bar{\Sigma}^{\text{xc}}
(\rr,\rr',\omega_i)-\bar{\Sigma}^{DC}(\rr,\rr',\omega_i),
\end{equation}
with a double counting correction $\bar{\Sigma}^{DC}(\rr,\rr',\omega_i)$ that is comprised of the SC-RDMFT exchange-correlation potentials:
\begin{equation}\label{selfenergy1}
\bar{\Sigma}^{\text{DC}, \sigma,\sigma'}(\rr,\rr',\omega_i)=\delta_{\sigma,\sigma'}\begin{bmatrix}
v_{\text{xc}}(\rr,\rr')&\Delta_{\text{xc}}(\rr,\rr')\\\Delta^*_{\text{xc}}(\rr,\rr')&-v_{\text{xc}}(\rr,\rr')
\end{bmatrix}.
\end{equation}

We now arrive at the reason 
for using the Kohn-Sham system as the non-interacting system of the Dyson 
equation. Indeed, the densities $(\gamma,\chi)$ can be written as the equal 
time limit of the Nambu-Gorkov Green's function. Here, $\gamma$ corresponds 
to the normal Green's function and $\chi$ to the anomalous propagator:
\begin{gather}
\gamma_{\sigma,\sigma'}(\rr,\rr')=\lim_{\eta \to 0^+} \frac{1}{\beta}\sum_{\omega_i}e^{i\eta \omega_i}G_{\sigma,\sigma'}(\rr,\rr',\omega_i)\\
\chi(\rr,\rr')=-\lim_{\eta \to 0^+} \frac{1}{\beta}\sum_{\omega_i}e^{i\eta \omega_i}F_{\uparrow,\downarrow}(\rr,\rr',-\omega_i).
\end{gather}
Obviously, this is true for the Nambu-Gorkov Kohn-Sham Green's function 
as well as for the Nambu-Gorkov Green's function of the interacting system. 
Since we defined the Kohn-Sham system as the non-interacting system 
reproducing $\chi$ and $\gamma$ of the interacting system, the equal time 
limit of $\bar{G}^{\text{KS}}(\rr\tau,\rr'\tau^+)=\bar{G}(\rr\tau,\rr'\tau^+)$. 
Inserting the equality into the corresponding Dyson equation yields the 
following two equations:

\begin{widetext}

\begin{gather}
0=\lim_{\eta \to 0^+} \frac{1}{\beta}\sum_{\omega_i,\sigma_1,\sigma_2}e^{i\eta \omega_i}\mdint{3}{r_1}{r_2}\Big[\bar{G}^{\text{KS}}_{\sigma,\sigma_1}(\rr,\rr_1,\omega_i)\bar{\Sigma}^{\text{KS}}_{\sigma_1\sigma_2}(\rr_1,\rr_2,\omega_i)\bar{G}_{\sigma_2,\sigma'}(\rr_2,\rr',\omega_i)\Big]_{11}\\
0=\lim_{\eta \to 0^+} \frac{1}{\beta}\sum_{\omega_i,\sigma_1,\sigma_2}e^{i\eta \omega_i}\mdint{3}{r_1}{r_2}\Big[\bar{G}^{\text{KS}}_{\sigma,\sigma_1}(\rr,\rr_1,-\omega_i)\bar{\Sigma}^{\text{KS}}_{\sigma_1\sigma_2}(\rr_1,\rr_2,-\omega_i)\bar{G}_{\sigma_2,\sigma'}(\rr_2,\rr',-\omega_i)\Big]_{12}.
\end{gather}
The analogue of this relation for the density is known for DFT 
\cite{ss1,ss2,ss3} as the Sham-Schl\"uter connection.
After decomposing the self-energy we arrive at the following system 
of integral equations for the exchange-correlation potentials. As a matter of notation we use
$\bar{G}^{\text{KS}}(\rr,\rr_1,\omega_i)=\bar{G}^{\text{KS}}_{\downarrow\downarrow}(\rr,\rr_1,\omega_i)=\bar{G}_{\uparrow\uparrow}^{\text{KS}}(\rr,\rr_1,\omega_i)$ 
and analogously for $F(\rr,\rr',\omega_i)$ and $\Sigma^{\text{xc}}(\rr,\rr',\omega_i)$.
\begin{gather}
\lim_{\eta \to 0^+} \frac{1}{\beta}\sum_{\omega_i,\sigma,\sigma_1,\sigma_2}e^{i\eta \omega_i}\mdint{3}{r_1}{r_2} \Big[\bar{G}^{\text{KS}}_{\sigma,\sigma_1}(\rr,\rr_1,\omega_i)\bar{\Sigma}^{\text{xc}}_{\sigma_1\sigma_2}(\rr_1,\rr_2,\omega_i)\bar{G}_{\sigma_2,\sigma}(\rr_2,\rr',\omega_i)\Big]_{11}\nonumber\\
=\lim_{\eta \to 0^+} \frac{2}{\beta}\sum_{\omega_i}\mdint{3}{r_1}{r_2} \Big[G^{\text{KS}}(\rr,\rr_1,\omega_i)v_{\text{xc}}(\rr_1,\rr_2)G(\rr_2,\rr',\omega_i)
-F^{\text{KS}}(\rr,\rr_1,\omega_i)v_{\text{xc}}(\rr_1,\rr_2)F^{\dagger}(\rr_2,\rr',\omega_i) \Big.\nonumber\\\Big.
-F^{\text{KS}}(\rr,\rr_1,\omega_i)\Delta_{\text{xc}}^*(\rr_1,\rr_2)G(\rr_2,\rr',\omega_i)
-G^{\text{KS}}(\rr,\rr_1,\omega_i)\Delta_{\text{xc}}(\rr_1,\rr_2)F^{\dagger}(\rr_2,\rr',\omega_i)\Big]
\end{gather}
and
\begin{gather}
\lim_{\eta \to 0^+} \frac{1}{\beta}\sum_{\omega_i,\sigma,\sigma_1,\sigma_2}e^{i\eta \omega_i}\mdint{3}{r_1}{r_2}\Big[\bar{G}^{\text{KS}}_{\sigma,\sigma_1}(\rr,\rr_1,-\omega_i)\bar{\Sigma}^{\text{xc}}_{\sigma_1\sigma_2}(\rr_1,\rr_2,-\omega_i)\bar{G}_{\sigma_2,\sigma}(\rr_2,\rr',-\omega_i)\Big]_{12}\nonumber\\
=-\lim_{\eta \to 0^+} \frac{2}{\beta}\sum_{\omega_i}\mdint{3}{r_1}{r_2} \Big[G^{\text{KS}}(\rr,\rr_1,-\omega_i)v_{\text{xc}}(\rr_1,\rr_2)F(\rr_2,\rr',-\omega_i)
+F^{\text{KS}}(\rr,\rr_1,-\omega_i)v_{\text{xc}}(\rr_1,\rr_2)G(\rr',\rr_2,\omega_i)\Big.\nonumber\\\Big.
-F^{\text{KS}}(\rr,\rr_1,-\omega_i)\Delta_{\text{xc}}^*(\rr_1,\rr_2)F(\rr_2,\rr',-\omega_i)+
G^{\text{KS}}(\rr,\rr_1,-\omega_i)\Delta_{\text{xc}}(\rr_1,\rr_2)G(\rr',\rr_2,\omega_i)\Big].
\end{gather}

One can slightly simplify these equations by transforming to Fourier-space
(the convergence factor $\lim_{\eta \to 0^+}$ is left out for simplicity's sake).
Indeed, by defining:
\begin{gather}
v_{\text{xc}}(\nk,\nk')=\mdint{3}{r_1}{r_2}\phi_k(\rr_2)\phi^*_{k'}(\rr_1) v_{\text{xc}}(\rr_1,\rr_2)
\\
\Sigma_{\text{xc}}(\nk,\nk',\omega_i)=\mdint{3}{r_1}{r_2} \phi_k(\rr_2)\phi^*_{k'}\Sigma_{\text{xc}}(\rr_1,\rr_2,\omega_i),
\end{gather}
the decoupling approximation reduces the anomalous potential to only 
one index $\nk$, namely:
\begin{gather}
\frac{1}{\beta}\sum_{\omega_i}\left[G^{\text{KS}}(\nk,\omega_i)\Sigma^{11}_{\text{xc}}(\nk,\nk',\omega_i)G(\nk',\omega_i)
-F^{\text{KS}}(\nk,\omega_i)\Sigma^{11}_{\text{xc}}(\nk,\nk',-\omega_i)F^{\dagger}(\nk',\omega_i) 
\right.\nonumber\\\left.
-F^{\text{KS}}(\nk,\omega_i)\Sigma_{\text{xc}}^{12*}(\nk,\nk',\omega_i)G(\nk',\omega_i)
-G^{\text{KS}}(\nk,\omega_i)\Sigma^{11}_{\text{xc}}(\nk,\nk',\omega_i)F^{\dagger}(\nk',\omega_i)\right]\nonumber\\
=
\frac{1}{\beta}\sum_{\omega_i} \left[G^{\text{KS}}(\nk,\omega_i)v_{\text{xc}}(\nk,\nk')G(\nk',\omega_i)
-F^{\text{KS}}(\nk,\omega_i)v_{\text{xc}}(\nk,\nk')F^{\dagger}(\nk',\omega_i) \right.\nonumber\\\left.
-\delta_{\nk,\nk'}\left(F^{\text{KS}}(\nk,\omega_i)\Delta_{\text{xc}}^*(\nk)G(\nk',\omega_i)+
G^{\text{KS}}(\nk,\omega_i)\Delta_{\text{xc}}(\nk)F^{\dagger}(\nk',\omega_i)\right)\right]\label{eq:53}\\
\frac{1}{\beta}\sum_{\omega_i}\left[G^{\text{KS}}(\nk,-\omega_i)\Sigma^{11}_{\text{xc}}(\nk,\nk',-\omega_i)F(\nk',-\omega_i)
-F^{\text{KS}}(\nk,-\omega_i)\Sigma^{11}_{\text{xc}}(\nk,\nk',\omega_i)G(\nk',\omega_i) 
\right.\nonumber\\\left.
+F^{\text{KS}}(\nk,-\omega_i)\Sigma_{\text{xc}}^{12*}(\nk,\nk',-\omega_i)F(\nk',-\omega_i)
+G^{\text{KS}}(\nk,-\omega_i)\Sigma^{11}_{\text{xc}}(\nk,\nk',\omega_i)G(\nk',\omega_i)\right]\nonumber\\
=\frac{1}{\beta}\sum_{\omega_i}\left[G^{\text{KS}}(\nk,-\omega_i)v_{\text{xc}}(\nk,\nk')F(\nk',-\omega_i)
+F^{\text{KS}}(\nk,-\omega_i)v_{\text{xc}}(\nk,\nk')G(\nk',\omega_i) 
\right.\nonumber\\\left.
-\delta_{\nk,\nk'}\left(F^{\text{KS}}(\nk,-\omega_i)\Delta_{\text{xc}}^*(\nk)F(\nk',-\omega_i)
-G^{\text{KS}}(\nk,-\omega_i)\Delta_{\text{xc}}(\nk)G(\nk',\omega_i)\right)\right].
\label{eq}
\end{gather}
\end{widetext}
At this point the first major advantage in comparison to 
SC-DFT emerges. In SC-DFT the equation corresponding to the 
density is of course local while the anomalous equation is 
non-local. The two equations arising from the Sham-Schl\"uter 
connection in SC-RDMFT are symmetric in the sense that they 
are both purely non-local. Consequently, we do not sum over 
any index $\nk$ in Eq.~\eqref{eq} and we are not plagued with 
the SC-DFT-problem of reconciling a local and a non-local equation. 
This greatly simplifies the calculations: the off-diagonal elements 
of $v_{\text{xc}}(\nk,\nk')$ do not depend on the anomalous potential 
and can be calculated separately. Moreover, 
if $v_{xc}$ satisfies the lattice periodicity (namely, 
$v_{xc}(\rr, \rr') = v_{xc}(\rr+\vecform{R}, \rr'+\vecform{R})$ with 
$\vecform{R}$ being any lattice vector), the nondiagonal 
elements are non-zero only when they belong to different bands
$v_{\text{xc}}(\nk,\vecform{n}'\vecform{k})$. Since the off-diagonal 
component $(\vecform{k}, \vecform{k}')$ becomes zero under the lattice
periodicity of the exchange-correlation potential, Eqs.~\eqref{eq:53} and 
\eqref{eq} can be further simplified.

In order to continue analogously to SC-DFT we simplify the equation by 
limiting ourselves to terms linear in $\Delta_{\nk}$:
\begin{gather}
\tilde{\Delta}_{\text{xc}}(\nk)=\sum_{\nk'}\frac{\delta \Delta_{\text{xc}}(\nk)}{\delta \Delta_{\nk'}}\Bigr|_{\substack{\Delta=0}}\Delta_{\nk'}.
\end{gather}
Naturally, this approximation is only valid close to the transition
temperature where the energy gap is small. As a last approximation, we
replace the interacting Green's function with the Kohn-Sham Green's
function.  From the perspective of Migdal's
theorem~\cite{migdal1958interaction}, this approximation is not
completely sound. For this reason, a promising avenue for further
research will be to dress the Green's function with the phononic self
energy~\cite{SCDFT2}.  The final form of the functional is derived in
detail in \hyperref[AppendixB]{Appendix B}.

We finally arrive at the linearized gap equation 
\begin{equation}
  \tilde{\Delta}_{\text{xc}}(\nk) = \tilde{\Delta}^A_{\text{xc}}(\nk) +
  \tilde{\Delta}^C_{\text{xc}}(\nk) + \tilde{\Delta}^D_{\text{xc}}(\nk),
\end{equation}
that we divide into one Coulomb term,
\begin{align}
\tilde{\Delta}^A_{\text{xc}}(\nk) & =-\frac{1}{2}\sum_{\nk'}\frac{\tanh\left(\frac{\beta\xi_{\nk'}}{2}\right)}{\xi_{\nk'}}\Delta_{\nk'} v(\nk,\nk'), 
\label{coulomb}
\end{align}
and two phonon terms
\begin{align}
\label{phonon1}
\tilde{\Delta}^C_{\text{xc}}(\nk) &=-\frac{1}{\tanh \left(\frac{\beta}{2}\xi_{\nk'}\right)}\sum_{\nk',\lambda q}\frac{\Delta_{\nk'}}{\xi_{\nk'}}\abs{g^{\nk\nk'}_{\lambda q}}^2
\\ &\qquad \nonumber
\Big[I(\xi_{\nk},\xi_{\nk'},\Omega_{\lambda q})-I(\xi_{\nk},-\xi_{\nk'},\Omega_{\lambda q})\Big],
\end{align}
and 
\begin{widetext}
\begin{align}
\tilde{\Delta}^D_{\text{xc}}(\nk)&=2\Delta_{\nk}\left(\frac{2\cosh^2\left(\frac{\beta}{2}\xi_{\nk}\right)}{\xi_{\nk}\beta}-\frac{\frac{\beta}{2}}{\sinh\left(\frac{\beta}{2}\xi_{\nk}\right)}\right)\sum_{\nk',\lambda q}\abs{g^{\nk\nk'}_{\lambda q}}^2I'(\xi_{\nk},\xi_{\nk'},\Omega_{\lambda q})
\nonumber\\
&-\frac{\Delta_{\nk}}{\tanh\left(\frac{\beta}{2}\xi_{\nk'}\right)}\sum_{\nk',\lambda q}\abs{g^{\nk\nk'}_{\lambda q}}^2\bigg\{\frac{1}{\xi_{\nk}}\Big[I(\xi_{\nk},\xi_{\nk'},\Omega_{\lambda q})-I(\xi_{\nk},-\xi_{\nk'},\Omega_{\lambda q})\Big]-2I'(\xi_{\nk},\xi_{\nk'},\Omega_{\lambda q})\bigg\}.
\label{phonon2}
\end{align}
\end{widetext}
The Coulomb term \eqref{coulomb} and phonon terms \eqref{phonon1} and
\eqref{phonon2} are completely analogous to the corresponding terms in
SC-DFT.  The only difference in the $A$ term lies in the fact that
the terms $\xi_{\nk}$, $\Delta_{\nk}$ and $g^{\nk\nk'}_{\lambda q}$
correspond to the results of RDMFT and not DFT calculations.  In
SCD-DFT there is an additional repulsive $B$ term, namely,
\begin{align*}
&\tilde{\Delta}^{B,{\rm DFT}}_{\text{xc}}(\nk)=
\frac{\Delta_{\nk}}{2}\left[\frac{1}{\xi_{\nk}}-\frac{\frac{\beta}{2}}{\cosh\left(\frac{\beta}{2}\xi_{\nk}\right)\sinh\left(\frac{\beta}{2}\xi_{\nk}\right)}\right]\nonumber\\
&\quad \times\Bigg\{\sum_{\nk'}\left[1-\tanh\left(\frac{\beta}{2}\xi_{\nk'}\right)\right]v(\nk,\nk')
- \\
&\quad \frac{\sum\limits_{\nk_1,\nk_2}\frac{\frac{\beta}{2}}{\cosh^2\left(\frac{\beta}{2}\xi_{\nk_1}\right)}\left[1-\tanh\left(\frac{\beta}{2}\xi_{\nk_2}\right)\right]v(\nk_1,\nk_2)}{\sum\limits_{\nk_1}\frac{\frac{\beta}{2}}{\cosh^2\left(\frac{\beta}{2}\xi_{\nk_1}\right)}}\Bigg\},
\end{align*}
which turns out to be rather problematic as it suddenly jumps to zero close 
to the Fermi surface and is consequently neglected~\cite{SCDFT1,SCDFT2}. 
Therefore, it speaks rather favorably of SC-RDMFT that the term is zero 
without any \textit{ad hoc} approximations. Once again the reason why the term is simpler 
in SC-RDMFT lies in the symmetry of the equations. The sums over $\nk_1$ in the last term arise due the use of the density in SC-DFT and consequently do not appear in SC-RDMFT.

The C-term \eqref{phonon1} is exactly analogous to superconducting DFT
term. The D-term \eqref{phonon2} is different from SC-DFT. 
The difference lies in the first summand which was neglected in SC-DFT as
it diverges \cite{SCDFT1}. Although the second summand also has a divergent 
contribution in SC-RDMFT, the divergences of the first and second summands 
asymptotically cancel  \cite{PhysRevB.88.014514}.
Consequently, the final form of the D-term is equal in both theories.
Comparing this functional
with its analogue from SC-DFT we only expect differences due to the
different Kohn-Sham energies and electron-phonon coupling in SC-RDMFT.
All in all, the process of developing the first anomalous
exchange-correlation functional for SC-RDMFT turned out to be simpler
and relied on fewer approximations to reach an equivalent result to
SC-DFT.

\section{Conclusions}
\label{conclusions} 

In this paper we presented the theoretical foundations of a new
\textit{ab initio} theory of superconductivity. SC-RDMFT is based on a
unique relationship between the statistical density operator at
equilibrium $\varrho_{\rm eq}$ and the corresponding one-body reduced
density matrix $\gamma$ and the anomalous density $\chi$. This reduced
density matrix formalism for superconductivity yields the existence of
a universal functional $\mathfrak{F}_\beta[\gamma, \chi]$ of these two
quantities, whose universal properties we derived.  The derived
theorems prove the possibility of a completely \textit{ab initio}
reduced density matrix formalism for superconductors and demonstrate
the elegance and advantages of using the Nambu-Gorkov one-particle
reduced density matrix or its substitutes to describe superconducting
systems.

We used a Bogoliubov transform to prove the existence of the Kohn-Sham system 
at finite temperature and obtain a system of coupled Bogoliubov-de Gennes-like 
Kohn-Sham equations. As even such coupled Kohn-Sham equations in their original form are
extremely challenging to solve we introduced the decoupling
approximation to SC-RDMFT. By decoupling the changes of
the superconducting phase transition from the electronic and phononic
calculations we were able to arrive at a linear system of
equations. Solving the system leads to a set of self-consistent
equations for the gap and our primary variables $\gamma$ and $\chi$.

We then used the existence of the Kohn-Sham system to derive an
exchange-correlation functional based on the Sham-Schl\"uter
connection in Nambu-Gorkov space. Formally the derivation was
extremely similar to SC-DFT but a few essential differences
existed. Because SC-RDMFT is concerned with two non-local variables,
the resulting equations exhibited a symmetry in their form that is
missing in SC-DFT. Due to this missing symmetry an extra strong
approximation is necessary in SC-DFT, namely that the system is nearly
homogeneous. In the end we simplified our functional to the linear
regime in order to arrive at a BCS-like gap equation. The resulting
approximate functional has three terms that have the same analytical form
as the analogous functional in SC-DFT. However, in SC-DFT there is a
fourth term that has a rather pathological behavior, that is set to
zero by hand. This problematic term turns out to be absent in SC-RDMFT.

Our functional was developed in close analogy to SC-DFT. It would be
certainly interesting to follow a different path, and to derive it by
generalizing existing (non-superconducting) RDMFT functionals (such as
the M\"uller functional). In view of the success of RDMFT for
correlated systems, this would maybe allow us to tackle, for the first
time, the problem of superconductivity in correlated systems in a
completely \textit{ab initio} fashion.

\acknowledgments
We thank Hardy Gross for helpful discussions.

\appendix

\begin{widetext}

\section{Nambu-Gorkov Green's function}
\label{AppendixA}

Just like the Nambu-Gorkov one-body reduced density matrix, 
the Green's function in Nambu-Gorkov space is a $2\times 2$ matrix:
\begin{gather}
\bar{G}_{\sigma,\sigma'}(\rr\tau,\rr'\tau')=
-\expval{\hat{T}\hat{\bar{\Psi}}(\rr,\tau)\otimes\hat{\bar{\Psi}}^\dagger(\rr',\tau')}
=\begin{bmatrix}
G_{\sigma,\sigma'}(\rr\tau,\rr'\tau') &\text{sign}(\sigma')F_{\sigma,-\sigma'}(\rr\tau,\rr'\tau')\\
\text{sign}(\sigma)F^\dagger_{-\sigma,\sigma'}(\rr\tau,\rr'\tau')&-\text{sign}(\sigma')\text{sign}(\sigma)G_{-\sigma,-\sigma'}(\rr\tau,\rr'\tau')
\end{bmatrix}.
\end{gather}
Here $G_{\sigma,\sigma'}(\rr\tau,\rr'\tau')$ corresponds to the normal 
Green's function and $F^\dagger_{-\sigma,\sigma'}(\rr\tau,\rr'\tau')$ 
is known as the anomalous propagator.

We can denote the constituents of the Kohn-Sham Green's function as:
\begin{gather}
G^{\text{KS}}_{\sigma,\sigma'}(\rr,\rr',\omega_i)=\delta_{\sigma,\sigma'}\sum_{\nk}\Big[\frac{u_{\nk}(\rr)u^*_{\nk}(\rr')}{i\omega_i-E_{\nk}}+\frac{v_{\nk}(\rr')v^*_{\nk}(\rr)}{i\omega_i+E_{\nk}}\Big]\\
F^{\text{KS}}_{\sigma,\sigma'}(\rr,\rr',\omega_i)=\delta_{\sigma,-\sigma'}\sum_{\nk}\Big[\frac{v_{\nk}(\rr')u^*_{\nk}(\rr)}{i\omega_i+E_{\nk}}-\frac{v_{\nk}(\rr)u^*_{\nk}(\rr')}{i\omega_i-E_{\nk}}\Big]
\end{gather}
in position space. A Fourier transformation assuming the decoupling 
approximation yields:
\begin{align}
G^{\text{KS}}_{\sigma,\sigma'}(\nk,\omega_i) &=
\mdint{3}{r}{r'} \phi^*_{\nk}(\rr')G^{\text{KS}}_{\sigma,\sigma'}(\rr,\rr',\omega_i)\phi_{\nk}(\rr)
=\delta_{\sigma,\sigma'}
\left(\frac{\abs{u_{\nk}}^2}{i\omega_i-E_{\nk}}+\frac{\abs{v_{\nk}}^2}{i\omega_i+E_{\nk}}\right),\\
F^{\text{KS}}_{\sigma,\sigma'}(\nk,\omega_i) &=
\mdint{3}{r}{r'} \phi^*_{\nk}(\rr')F^{\text{KS}}_{\sigma,\sigma'}(\rr,\rr',\omega_i)\phi_{\nk}(\rr)
=\delta_{\sigma,-\sigma'}\text{sign}(\sigma')u_{\nk}v^*_{\nk}\left(\frac{1}{i\omega_i+E_{\nk}}-\frac{1}{i\omega_i-E_{\nk}}\right).
\end{align}

\section{Linear Regime and Frequency Sums}
\label{AppendixB}

In the linear regime we arrive at:
\begin{gather}\label{eq55}
\tilde{\Delta}_{\text{xc}}(\nk)\frac{1}{\beta}\sum_{\omega_i}\tilde{G}^{\text{KS}}(\nk,-\omega_i)\tilde{G}(\nk,\omega_i)\frac{1}{\beta}\sum_{\omega_i}\tilde{G}^{\text{KS}}(\nk,\omega_i)\tilde{G}(\nk,\omega_i)
\nonumber\\
=\frac{1}{\beta}\sum_{\omega_i}\tilde{G}^{\text{KS}}(\nk,\omega_i)\tilde{G}(\nk,\omega_i) \frac{1}{\beta}\sum_{\omega_i}\Bigg\{\tilde{G}^{\text{KS}}(\nk,-\omega_i)\tilde{\Sigma}^{11}_{\text{xc}}(\nk,\nk,-\omega_i)\tilde{F}(\nk,-\omega_i)\Bigg.\nonumber\\\Bigg.+\tilde{F}^{\text{KS}}(\nk,-\omega_i)\tilde{\Sigma}^{11}_{\text{xc}}(\nk,\nk,\omega_i)\tilde{G}(\nk,\omega_i)+\tilde{G}^{\text{KS}}(\nk,-\omega_i)\tilde{\Sigma}^{12}_{\text{xc}}(\nk,\nk,-\omega_i)\tilde{G}(\nk,\omega_i)\Bigg\}\nonumber\\ 
-\frac{1}{\beta} \sum_{\omega_i}\tilde{G}^{\text{KS}}(\nk,\omega_i)\tilde{\Sigma}^{11}_{\text{xc}}(\nk,\nk,\omega_i)\tilde{G}^{\text{KS}}(\nk,\omega_i)
\times\sum_{\omega_i}\Bigg[\tilde{G}^{\text{KS}}(\nk,-\omega_i)\tilde{F}(\nk,-\omega_i)+\tilde{F}^{\text{KS}}(\nk,-\omega_i)\tilde{G}(\nk,\omega_i)\Bigg].
\end{gather}

Fortunately, Marques~\cite{marquesthesis} already calculated all the required frequency sums analytically 
\footnote{$\tilde{\epsilon}_{\nk}$ is replaced with $\xi_{\nk}$ to avoid any confusion with the other variables in the linear regime}:
\begin{gather}
\frac{1}{\beta}\sum_{\omega_i}\big[\tilde{G}^{\text{KS}}(\nk,\omega_i)\tilde{G}^{\text{KS}}(\nk,-\omega_i)\big]=\frac{1}{\beta}\sum_{\omega_i}\frac{1}{i\omega_i-\xi_k}\frac{1}{-i\omega_i-\xi_k}=\frac{1}{2\xi_{\nk}}\tanh\left(\frac{\beta}{2}\xi_{\nk}\right)
\end{gather}
\begin{gather}
\frac{1}{\beta}\sum_{\omega_i}\big[\tilde{G}^{\text{KS}}(\nk,\omega_i)\tilde{G}^{\text{KS}}(\nk,\omega_i)\big]=\frac{1}{\beta}\sum_{\omega_i}\frac{1}{i\omega_i-\xi_k}\frac{1}{-i\omega_i-\xi_k}=-\frac{1}{2}\frac{\frac{\beta}{2}}{\cosh^2\left(\frac{\beta}{2}\xi_{\nk}\right)}
\end{gather}
\begin{gather}
\frac{1}{\beta}\sum_{\omega_i}\tilde{F}^{\text{KS}}(\nk,\omega_i)\big[\tilde{G}^{\text{KS}}(\nk,\omega_i)+\tilde{G}^{\text{KS}}(\nk,-\omega_i)\big]=
\frac{1}{\beta}\sum_{\omega_i}\frac{1}{i\omega_i-\xi_k}\frac{\Delta_k}{\omega_i^2+\xi_k^2}\nonumber\\=\frac{d}{d\xi_k}\frac{1}{\beta}\sum_{\omega_i}\frac{\Delta_k}{\omega_i^2+\xi_k^2}=\frac{d}{d\xi_k}\left(\frac{1}{2\xi_k}-\frac{1}{\xi_k}\frac{1}{1+e^{\beta \xi_k}}\right)=\frac{\Delta_{\nk}}{2\xi_{\nk}}\left[\frac{\frac{\beta}{2}}{\cosh^2\left(\frac{\beta}{2}\xi_{\nk}\right)}-\frac{\tanh\left(\frac{\beta}{2}\xi_{\nk}\right)}{\xi_{\nk}}\right]
\end{gather}
For the electronic self-energy expressions we arrive at:
\begin{gather}
\frac{1}{\beta}\sum_{\omega_i}\tilde{G}^{\text{KS}}(\nk,\omega_i)\tilde{\Sigma}^{11}_{\text{xc}}(\nk,\nk,\omega_i)\tilde{G}^{\text{KS}}(\nk,\omega_i)
=\frac{1}{4}\sum_{\nk'}
\frac{\frac{\beta}{2}}{\cosh^2\left(\frac{\beta}{2}\xi_{\nk}\right)}\left[1-\tanh\left(\frac{\beta}{2}\xi_{\nk'}\right)\right]v(\nk,\nk')
\end{gather}
\begin{gather}
\frac{1}{\beta}\sum_{\omega_i}\tilde{G}^{\text{KS}}(\nk,\omega_i)\tilde{\Sigma}^{12}_{\text{xc}}(\nk,\nk,\omega_i)\tilde{G}^{\text{KS}}(\nk,-\omega_i)
=-\frac{1}{4}\sum_{\nk'}
\frac{\Delta_{\nk'}}{\xi_{\nk}\xi_{\nk'}}\tanh\left(\frac{\beta}{2}\xi_{\nk}\right)\tanh\left(\frac{\beta}{2}\xi_{\nk'}\right)v(\nk,\nk')
\end{gather}
\begin{gather}
\frac{1}{\beta}\sum_{\omega_i}\tilde{G}^{\text{KS}}(\nk,\omega_i)\tilde{\Sigma}^{11}_{\text{xc}}(\nk,\nk,\omega_i)\tilde{F}^{\text{KS}}(\nk,\omega_i)\nonumber\\
=\frac{1}{8}\frac{\Delta_{\nk}}{\xi_{\nk}}\left[\frac{1}{\xi_{\nk}}\tanh\left(\frac{\beta}{2}\xi_{\nk}\right)-\frac{\frac{\beta}{2}}{\cosh^2\left(\frac{\beta}{2}\xi_{\nk}\right)}\right]\times \sum_{\nk'}\left[1-\tanh\left(\frac{\beta}{2}\xi_{\nk'}\right)\right]v(\nk,\nk').
\end{gather}
And the phononic terms yield:
\begin{gather}
\frac{1}{\beta}\sum_{\omega_i}\tilde{G}^{\text{KS}}(\nk,\omega_i)\tilde{\Sigma}^{11}_{\text{xc}}(\nk,\nk,\omega_i)\tilde{G}^{\text{KS}}(\nk,\omega_i)
=-\sum_{\nk',\lambda q}\abs{g^{\nk\nk'}_{\lambda q}}^2I'(\xi_{\nk},\xi_{\nk'},\Omega_{\lambda q})
\end{gather}
\begin{gather}
\frac{1}{\beta}\sum_{\omega_i}\tilde{G}^{\text{KS}}(\nk,\omega_i)\tilde{\Sigma}^{12}_{\text{xc}}(\nk,\nk,\omega_i)\tilde{G}^{\text{KS}}(\nk,-\omega_i)=-\frac{1}{2}\sum_{\nk',\lambda q}\frac{\Delta_{\nk'}}{\xi_{\nk}\xi_{\nk'}}\abs{g^{\nk\nk'}_{\lambda q}}^2\Big[I(\xi_{\nk},\xi_{\nk'},\Omega_{\lambda q})-I(\xi_{\nk},-\xi_{\nk'},\Omega_{\lambda q})\Big]
\end{gather}
\begin{gather}
\frac{1}{\beta}\sum_{\omega_i}\tilde{G}^{\text{KS}}(\nk,\omega_i)\tilde{\Sigma}^{11}_{\text{xc}}(\nk,\nk,\omega_i)\tilde{F}^{\text{KS}}(\nk,\omega_i)
=\frac{1}{\beta}\sum_{\omega_i}\tilde{G}^{\text{KS}}(\nk,-\omega_i)\tilde{\Sigma}^{11}_{\text{xc}}(\nk,\nk,-\omega_i)\tilde{F}^{\text{KS}}(\nk,\omega_i)\nonumber\\
=\frac{\Delta_{\nk}}{2\xi_{\nk}}\sum_{\nk',\lambda q}\abs{g^{\nk\nk'}_{\lambda q}}^2\bigg\{I'(\xi_{\nk},\xi_{\nk'},\Omega_{\lambda q})-\frac{1}{2\xi_{\nk}}\Big[I(\xi_{\nk},\xi_{\nk'},\Omega_{\lambda q})-I(\xi_{\nk},-\xi_{\nk'},\Omega_{\lambda q})\Big]\bigg\}
\end{gather}
with
\begin{gather}
I(E,E',\Omega_{\lambda q})=\frac{1}{\beta}\sum_{\omega_1,\omega_2}G^{\text{KS}}(E,\omega_i)D_{\lambda q}(\omega_1-\omega_2)G^{\text{KS}}(E',\omega_2)
\end{gather}
and $D_{\lambda q}(\omega_1-\omega_2)$ being the phonon propagator.

\end{widetext}

\bibliography{Ares}

\end{document}